\documentclass[aps,prb,floatfix,amsmath,amssymb,superscriptaddress,twocolumn,eqsecnum,nofootinbib]{revtex4-2}
\usepackage{bm}                     
\usepackage{appendix}
\usepackage[latin1]{inputenc}
\usepackage{dcolumn}      
\usepackage{bbm}
\usepackage{color}
\usepackage{xcolor}
\usepackage{amssymb}
\usepackage{amsmath}
\usepackage[mathscr]{eucal}
\usepackage{latexsym}
\usepackage{amsbsy}
\usepackage{float}
\usepackage{graphicx}
\usepackage{epsfig}
\usepackage{subfigure}
\usepackage{wrapfig}
\usepackage{epsf}
\usepackage{float}
\usepackage[normalem]{ulem}

\newcommand{\ba}{\begin{array}}
\newcommand{\ea}{\end{array}}
\newcommand{\be}{\begin{equation}}
\newcommand{\ee}{\end{equation}}
\newcommand{\bea}{\begin{eqnarray}}
\newcommand{\eea}{\end{eqnarray}}

\begin{document}

\title{Two-dimensional Dirac fermions in a mass superlattice}

\author{Alessandro De Martino}
\affiliation{Department of Mathematics, City, University of London, 
London EC1V 0HB, United Kingdom}

\author{Luca Dell'Anna}
\affiliation{Dipartimento di Fisica e Astronomia G. Galilei, Universit\`a degli studi di Padova, 
35131 Padova, Italy}

\author{Lukas Handt}
\affiliation{Institut f\"ur Theoretische Physik,
Heinrich-Heine-Universit\"at, D-40225  D\"usseldorf, Germany}

\author{Andrea Miserocchi}
\affiliation{Dipartimento di Fisica e Astronomia G. Galilei, Universit\`a degli studi di Padova, 
35131 Padova, Italy}
\affiliation{Institut f\"ur Theoretische Physik,
Heinrich-Heine-Universit\"at, D-40225  D\"usseldorf, Germany}

\author{Reinhold Egger}
\affiliation{Institut f\"ur Theoretische Physik,
Heinrich-Heine-Universit\"at, D-40225  D\"usseldorf, Germany}

\begin{abstract}
We study two-dimensional (2D) Dirac fermions in the presence of 
a periodic mass term alternating between positive and negative values
along one direction.  This scenario could be realized for a graphene monolayer or 
for the surface states of topological insulators. 
The low-energy physics is governed by chiral Jackiw-Rebbi modes propagating along zero-mass lines, 
with the energy dispersion of the Bloch states given by an anisotropic Dirac cone.
By means of the transfer matrix approach, we obtain exact results for a piece-wise constant
mass superlattice. On top of Bloch states, two different classes of boundary and/or 
interface modes can exist in a finite-size geometry or 
in a nonuniform electrostatic potential, respectively.  
We compute the dispersion relation for both types of boundary and interface modes,
which originate either from states close to the superlattice Brillouin zone (BZ) center or, via a Lifshitz transition, 
from states near the BZ boundary. In the presence of a potential step, we predict that the interface modes, the 
Bloch wave functions, and the electrical conductance will sensitively depend on the step position relative 
to the mass superlattice.
\end{abstract}
\date{\today}

\maketitle

\section{Introduction}\label{sec1}

It is well known that the band structure of solids can be modified in a controllable way by 
means of superlattice potentials. For instance, the use of electrostatic superlattice potentials 
has been suggested as versatile and tunable tool for creating emergent Dirac fermions 
with anisotropic dispersion in 2D graphene monolayers \cite{park2008,park2008b,Isacsson2008,Barbier2010,Ponomarenko2013,
Forsythe2018,Li2021c} 
or in few-layer black phosphorus devices \cite{Li2017a}.  
Similarly, moir{\'e} superlattice effects can induce a spectacular 
restructuring of the band structure in twisted bilayer graphene \cite{Ledwith2021},
layered van der Waals materials \cite{Ideue2021}, and topological insulators (TIs) \cite{Cano2021}, 
including the formation of topologically nontrivial and nearly flat bands with strong correlation effects \cite{Andrei2020}.   
Apart from the mostly considered case of electrostatic superlattices, 
interesting modifications of the band structure have also been predicted for magnetic superlattices
and for periodic modulations of the spin-orbit coupling, 
see, e.g., Refs.~\cite{DellAnna2009,Tan2010,DellAnna2011,Lenz2011} for the case of graphene monolayers.
 
In the present work, we focus on yet another superlattice type which can be realized 
in 2D Dirac materials, e.g., in  graphene monolayers \cite{GrapheneReview2009} or 
the surface states of TIs \cite{Hasan2010,Shen2017}.
We study the effects of a one-dimensional (1D) \emph{mass superlattice} $M(x)$, 
which periodically alternates between regions of positive and negative mass. 
(The mass term is assumed homogeneous along the $y$-direction, with the 2D material in the $xy$-plane.) 
For the graphene  case, such a mass profile could arise
from a sublattice-dependent  potential due to substrate or strain effects \cite{GrapheneReview2009}. 
For TI surface states, it could (approximately) be generated by the exchange field of 
an array of magnetic stripes with alternating magnetization direction.  

It is well known that a single mass kink binds a fermionic zero mode 
by the Jackiw-Rebbi mechanism \cite{JackiwRebbi1976,Jackiw1981,Semenoff2008}. 
This zero mode is unidirectional (``chiral'') and propagates
with the Fermi velocity $v_{\rm F}$ either in the positive or negative $y$-direction 
while being exponentially localized near the mass kink along the $x$-direction.  
In general terms, a sign change of the mass for 2D Dirac fermions corresponds to
a transition between two topological Chern insulators with a different Chern number \cite{QAHreview}. 
By the bulk-boundary correspondence,  zero-mass lines at the interfaces then harbor chiral zero modes.
For the TI realization, experimental evidence for such chiral zero modes has been reported 
in Refs.~\cite{yasuda2017,Rosen2017}.  
In Bernal-stacked bilayer graphene devices, in the presence of either interlayer bias voltage kinks, tilt boundaries, or  in folded geometries, one expects topological valley-momentum-locked zero-line modes \cite{Martin2008,Qiao2014} that closely resemble 
the above chiral  zero mode \cite{Yao2009,Bi2015}. We refer the reader to
Ref.~\cite{Wang2021a} for a recent survey, including a summary of the
 experimental evidence for zero-line modes in bilayer graphene.
In particular, such modes have been identified by scanning tunneling microscopy (STM) \cite{Yin2016}.  
Similar zero-line modes also appear in the helical network description of minimally twisted bilayer graphene \cite{Prada2012}.
More generally, depending on the symmetries of the problem, 1D zero-line modes can 
also appear near line defects such as dislocations \cite{Ran2009,Teo2010}.

For 2D Dirac fermions with  a periodic mass $M(x)$ alternating between positive and negative values,
chiral 1D modes are located near the positions with $M(x)=0$,  with adjacent modes having 
opposite propagation direction. 
While low-energy transport remains efficient 
along the $y$-direction, the band structure flattens along the $x$-direction.
For large mass amplitude (and assuming the same absolute value for positive and negative mass regions), 
the residual overlap between counterpropagating neighboring chiral modes generates a
small velocity $v_x\ll v_{\rm F}$ along the $x$-direction. In effect, one then arrives at a highly anisotropic 
Dirac cone dispersion at low energies  \cite{Zhou2018a,Xiao2020}.
We here show that the case of a piece-wise constant periodic mass term is exactly
solvable.  Our calculations confirm the existence of anisotropic Dirac cones,
yield analytical results for the ratio $v_x/v_{\rm F}$, and 
provide a useful starting point for future studies of interaction effects and/or magnetic fields.
We note that in Refs.~\cite{zarenia2012,maksimova2012}, closely related models have been studied.
In particular, the authors of Ref.~\cite{zarenia2012} show that for smooth mass kinks, 
additional non-chiral localized states 
analogous to  Volkov-Pankratov states \cite{Goerbig2017,tineke2020} can exist.  
However, the anisotropy of the Dirac cone dispersion has not been discussed 
in Ref.~\cite{zarenia2012}.
Moreover, while Ref.~\cite{maksimova2012} (see also Ref.~\cite{BrionesTorres2014}) contains a 
detailed discussion of the electronic spectrum for a periodic mass problem, their mass term 
alternates between zero and a finite value, in contrast to the mass term considered
below.  As a consequence, chiral zero modes and physical effects caused by these modes are absent 
in Refs.~\cite{maksimova2012,BrionesTorres2014}.  Let us also mention that
we here study a coupled-wire model, see Refs.~\cite{Vishwanath2001,Kane2002} for related but different examples, 
where the 1D wires correspond to chiral zero modes with alternating propagation direction \cite{Han2022}. 

A central result of our work is to point out the existence of \emph{two 
types of boundary modes} in the presence of a sample boundary along the $y$-direction.
The modes are spatially confined to the vicinity of the boundary but can propagate along the
boundary.
Similarly, for an electrostatic potential step along the $x$-direction, 
we predict two types of \emph{interface modes}.
The two different mode types emerge either near the center of the superlattice BZ or  
near the BZ boundary. In the latter case, we observe that such modes appear only
if the mass amplitude exceeds a critical value.  Under this condition, the Fermi surface for 
the lowest band undergoes a Lifshitz transition \cite{Lifshitz1960}, opening
up from a closed elliptic contour into a pair of open (disconnected) arcs.
Remarkably,  
both types of boundary and/or interface modes can only exist 
in the presence of the mass superlattice, and their spatial decay length can exceed the 
lattice constant of the mass term.  

The structure  of this paper is as follows. In Sec.~\ref{sec2}, we introduce  
the model and the assumptions behind it, and we consider the cases of a 
single mass kink and of a mass barrier.  (Technical details have been delegated to the Appendix.) 
Next, in Sec.~\ref{sec3} we use the transfer matrix approach to determine 
the band structure and the Bloch states for a 
piecewise periodic mass term with alternating regions of mass $\pm M$, see Eq.~\eqref{periodicmass} below.  
In this case, we find a gapless low-energy anisotropic Dirac cone near the $\Gamma$ point of the superlattice BZ.
However,  if
the positive and negative mass amplitudes differ, a spectral gap will open, as shown in Sec.~\ref{sec3c},
where we construct a systematic low-energy theory.
Importantly, in the presence of boundaries or in an inhomogeneous electrostatic potential,
the spectral condition also allows for evanescent wave solutions.  We discuss 
 boundary modes in Sec.~\ref{sec4}.  In Sec.~\ref{sec5}, we include  
 an electrostatic potential step along the $x$-direction, 
 which defines an $np$-junction. We determine the transmission probability for 
 Bloch states and show that the conductance across the step will 
 sensitively depend on the step position. This dependence is a direct consequence
 of the fact that low-energy states have significant weight only near the positions of mass (anti-)kinks.
 In Sec.~\ref{sec5c}, we show that interface modes of various types can exist and we compute
 their energy dispersion. The paper concludes with an outlook in Sec.~\ref{sec6}.
 
\section{Model}\label{sec2}

In this paper, we study noninteracting electrons described by a 2D Dirac Hamiltonian with a single Dirac cone.  
This model captures the essential physics of the spin-momentum locked and protected surface states in 3D  TI materials \cite{Hasan2010,Shen2017}, 
as well as the low-energy physics of 2D graphene monolayers which is governed by states close to a single $K$ point (``valley'') \cite{GrapheneReview2009}.  
For the latter case, the assumption of a single $K$ point requires   
the mass or potential terms considered below to be actually smooth on the scale of the lattice spacing of graphene.
For an infinitely extended system in the $xy$-plane, using units with $\hbar=1$ and Fermi velocity $v_{\rm F}=1$ throughout, we study the Hamiltonian 
\be\label{ham}
H=-i\sigma_x\partial_x-i\sigma_y \partial_y + M(x)\sigma_z +V(x) \mathbbm{1},
\ee
with the electrostatic potential $V(x)$ and the mass term $M(x)$. Both terms 
are assumed homogeneous along the $y$-direction.  As a consequence of this translation invariance, 
the wave vector (or momentum) component $k_y$ is conserved.  The Pauli matrices $\sigma_{x,y,z}$ and the $2\times 2$ identity matrix $\mathbbm{1}$ act in spin space for TI surface states, and in the 
sublattice space of the honeycomb lattice for the case of graphene. 

For given momentum $k_y$, the spinor eigenstates of Eq.~\eqref{ham} can be written as  
\be\label{spinor}
\Psi(x,y)= e^{ik_yy}\, \psi(x),\quad \psi(x) =\begin{pmatrix} u(x) \\ v(x)\end{pmatrix},
\ee
which results in the 1D Dirac equation 
\be
\label{Dirac}
\begin{pmatrix}   M(x) +V(x)& -i(\partial_x+k_y)\\
  -i(\partial_x -k_y) & -M(x) +V(x)\end{pmatrix}\begin{pmatrix}
u \\ v\end{pmatrix}=E\begin{pmatrix}u \\ v\end{pmatrix}.
\ee
In this work, we are interested in the case of a spatially periodic mass term which alternates between positive and negative values. 
As simple and exactly solvable model, we will consider the piece-wise constant periodic mass term discussed in Sec.~\ref{sec3}. 
For the TI case, such a mass term can (approximately) be generated by 
the deposition of ferromagnetic insulator stripes with alternating magnetization on a TI surface, where the magnetic exchange contributions produce a periodic mass term \cite{Xiao2020}.
Similarly, for a graphene monolayer, a suitably patterned substrate creates
a sublattice-dependent superlattice potential which in effect gives a periodic mass term \cite{GrapheneReview2009}.

In the remainder of this section, to prepare the ground for the 
periodic mass case in Sec.~\ref{sec3}, we will analyze three simpler problems. 
In Sec.~\ref{sec2a},  we determine the general solution of Eq.~\eqref{Dirac} for the 
homogeneous case. In Sec.~\ref{sec2b}, we rederive the well-known
 low-energy spectrum for a mass kink, $M(x)=M \, {\rm sgn}(x)$,
which binds a 1D chiral zero mode propagating along the $y$-direction 
\cite{JackiwRebbi1976,Jackiw1981,Semenoff2008,Hasan2010}.
In Sec.~\ref{sec2c}, we study a mass barrier composed of a mass kink and an anti-kink,
where one finds two counterpropagating chiral zero modes. 
For ease of notation, we often keep the dependence on $k_y$ and $E$ implicit. 

\subsection{Homogeneous problem}\label{sec2a}

Let us first specify the general (not normalized) eigenstates of Eq.~\eqref{Dirac} for a region with 
constant potential, $V(x)=V$, and constant mass, $M(x)=M$.  A uniform scalar potential can be included by shifting $E\to E-V$, which we implicitly assume below. 
For $M(x)=M$, the solution is given by
\be\label{W}
\psi(x) =  W_{M}(x) \begin{pmatrix} a \\ b\end{pmatrix},
\ee
where $a$ and $b$ are arbitrary complex coefficients and we define the matrix
\be\label{WM}
W_{M}(x) = \begin{pmatrix}
e^{\kappa x} & e^{-\kappa x} \\
i\frac{ k_y-\kappa}{M+E} e^{\kappa x}  &i \frac{k_y+\kappa}{M+E} e^{-\kappa x}  
\end{pmatrix},  
\ee
with the definition
\be\label{kappadef}
\kappa= \left\{\begin{array}{cc} \sqrt{M^2+k_y^2 -E^2}, & E^2< k_y^2+M^2,\\
ik\equiv i\sqrt{E^2-M^2-k_y^2}, & E^2> k_y^2+M^2. \end{array}\right.
\ee
For low energies, $E^2<k_y^2+M^2$, we have evanescent waves along the $x$-direction, 
and the eigenstates are spatially localized on the 
length scale $\kappa^{-1}$ near boundaries or mass kinks.
For $E^2>k_y^2+M^2$, $\kappa=ik$ is purely imaginary and we find plane-wave 
solutions propagating along the $x$-direction with wave number $k_x=k$.
Useful expressions involving $W_{M}(x)$ 
in Eq.~\eqref{WM} are summarized in Appendix~\ref{appA}.
In particular, Eqs.~\eqref{jxx1} and \eqref{jxx2} imply that the $x$-component of the 
particle current density is given by
\be\label{jx}
j_x = \psi^\dagger \sigma_x \psi = \left\{  
\begin{array}{cl}
 \frac{4\kappa {\rm Im}(b^*a)}{M+E}, &E^2< k_y^2+M^2, \\
 & \\
\frac{2k(|a|^2-|b|^2)}{M+E}, & E^2> k_y^2+M^2.
\end{array}
\right. 
\ee

\subsection{Mass kink}\label{sec2b}

We turn to the case of a single mass kink, $M(x)=M\, {\rm sgn}(x)$ with $M>0$, see Ref.~\cite{Semenoff2008}.
We here discuss only the low-energy case, $E^2< k_y^2+M^2$, where $\kappa$ in Eq.~\eqref{kappadef} is real.
From Eq.~\eqref{W}, normalizable eigenstates then have the form
\be\label{ansatz}
\psi(x) = \left\{
\begin{array}{cc}
  W_{-M}(x) \begin{pmatrix} 
a_L \\ 0
\end{pmatrix}   &  \text{for}\; x<0, \\
W_{M}(x) \begin{pmatrix} 
0 \\ b_R
\end{pmatrix}    &  \text{for} \;  x>0, 
\end{array} \right.
\ee
where the coefficients $a_L$ and $b_R$ are determined by continuity of 
$\psi(x)$ at $x=0$ and normalization.  
Using Eq.~\eqref{aux1}, we define the matrix 
\bea\label{Om1}
\Omega_M&=&W^{-1}_{M}(0) \,W_{-M}^{}(0)  \\ 
\nonumber &=&\frac{1}{\kappa(E-M)}
\begin{pmatrix}
E \kappa - k_y M  &  -(\kappa + k_y) M \\
(-\kappa + k_y) M & E \kappa + k_y M
\end{pmatrix},
\eea
such that the continuity condition takes the form
\be\label{cont1}
\begin{pmatrix} 0 \\ b_R\end{pmatrix} = \Omega_M \begin{pmatrix} 
a_L \\ 0\end{pmatrix}.
\ee
As a result, we get the relations
$0=(E\kappa - k_y M)a_L$ and  $b_R =\frac{(-\kappa + k_y) M}{\kappa(E-M)} a_L$. 
For nontrivial solutions, we must have $E\kappa-k_yM=0$ from the first relation, which is solved by 
the dispersion relation $E(k_y)=k_y$ of a 1D chiral mode. 
The second relation then yields $b_R=a_L$ for the spinor wave function, 
where $a_L$ is finally determined by normalization.  
This chiral mode propagates with  velocity $v_{\rm F}$ along the positive $y$-direction
and is localized near the mass kink at $x=0$ in the $x$-direction. 
Similarly, for an anti-kink mass profile with $M$ replaced by $-M$, 
one finds a 1D chiral mode propagating along the negative $y$-direction, 
with dispersion relation $E(k_y)=-k_y$.

\subsection{Mass barrier}\label{sec2c}

Next we consider a mass barrier of width $\ell$ described by~\cite{Qiao2014}  
\be
\label{massbarrier}
M(x) = \left\{ \begin{array}{cc}
  M   & \textrm{for} \; |x|< \ell/2, \\
 -M   &  \textrm{for} \; |x|>\ell/2.
\end{array}
\right. 
\ee
We search for low-energy solutions with $E^2<k_y^2+ M^2$, where
normalizable eigenstates can be written as
\be
\psi(x) = \left\{\begin{array}{ll} W_{-M}(x) \begin{pmatrix} 
a_L \\ 0
\end{pmatrix}   &  \text{for} \; x<-\ell/2, \\
W_{M}(x) \begin{pmatrix} 
a \\ b
\end{pmatrix}   &  \text{for} \; |x|<\ell/2, \\
W_{-M}(x) \begin{pmatrix} 
0 \\ b_R
\end{pmatrix}    &  \text{for} \;  x>\ell/2 ,
\end{array} \right.
\ee
with coefficients $a_L, a, b,$ and $b_R$. 
Imposing  continuity at $x=\pm \ell/2$, one can eliminate $a$ and $b$. We arrive at Eq.~\eqref{cont1} but with $\Omega_M$ replaced by
\be
\Omega_B = W^{-1}_{-M}(\ell/2)\, W^{}_{M}(\ell/2) \, W^{-1}_{M}(-\ell/2) \,
W_{-M}^{}(-\ell/2),\label{OmegaB}
\ee
see Eq.~\eqref{OmegaBA} for explicit matrix elements.
The dispersion relation follows from $\left[ \Omega_{B}\right]_{11}=0$, 
which reads explicitly
\be\label{eqbarrier}
E^2 = k_y^2 + M^2 e^{-2\kappa \ell}.
\ee
For barrier width $\ell\to\infty$, we can neglect the exponential term  and obtain  $E_\pm(k_y)=\pm k_y$,
corresponding to a pair of counterpropagating chiral zero modes localized at the barrier edges. 
For large but finite barrier width with $M\ell\gg 1$, the two chiral zero modes hybridize. 
The level crossing at $k_y=0$ is now replaced by an avoided crossing, where 
Eq.~\eqref{eqbarrier} yields $E_\pm(k_y=0) \simeq \pm Me^{-\ell M}.$
The low-energy dispersion then acquires an exponentially 
small gap due to the avoided crossing,
$E_\pm(k_y) \simeq \pm \sqrt{k_y^2 + M^2 e^{-2M \ell}}$.

\section{Periodic mass}\label{sec3}

\begin{figure}[t]
\centering
\includegraphics[width=\columnwidth]{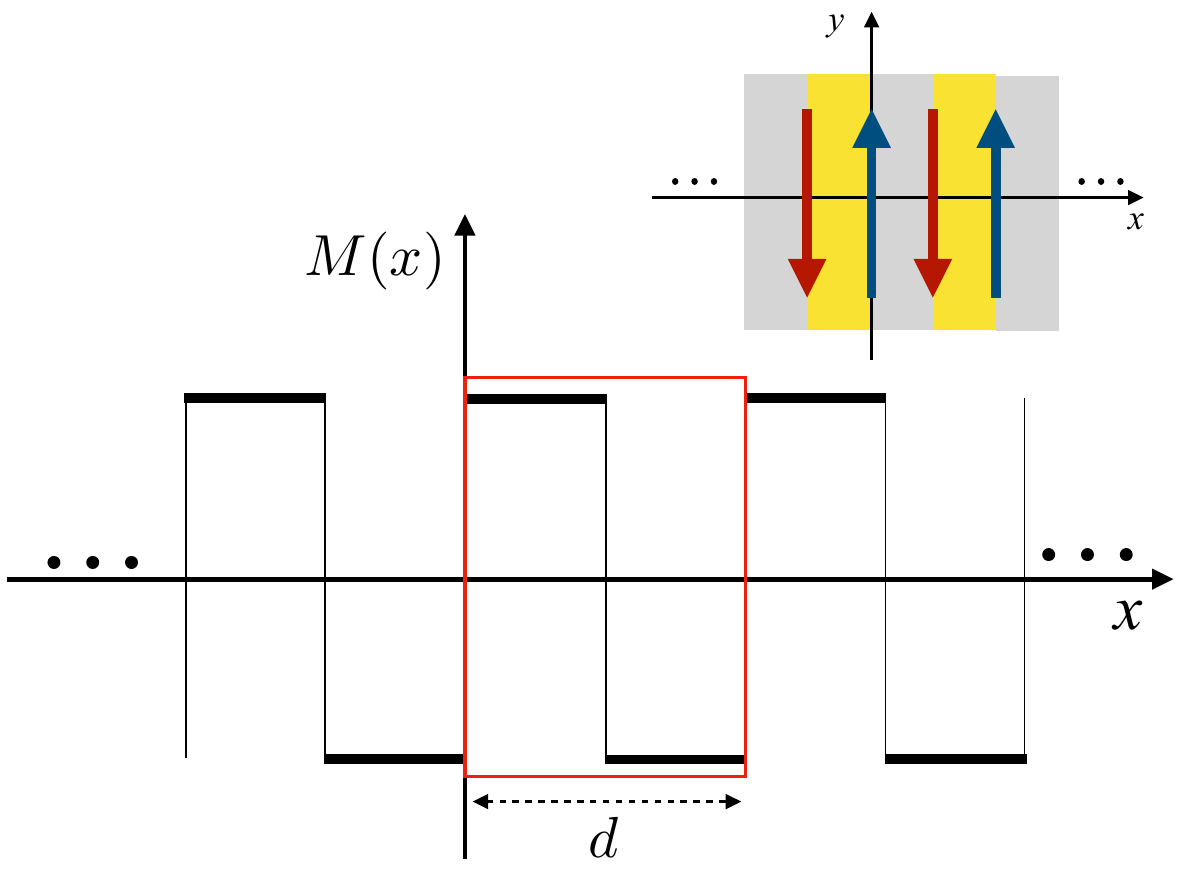}
\caption{Piece-wise constant periodic mass profile $M(x)$ in Eq.~\eqref{periodicmass}. 
A unit cell of length $d$ is indicated by the red square. The inset 
indicates the regions of positive (grey) and negative (yellow) mass in the $xy$-plane.  
1D chiral zero modes are generated near the (anti-)kink positions 
by the Jackiw-Rebbi mechanism, with the respective propagation direction indicated by arrows. }
\label{fig1}
\end{figure}

In this section, we discuss the solution of the Dirac equation \eqref{Dirac} for the
 piece-wise constant periodic mass term sketched in Fig.~\ref{fig1}, which is given by 
\be\label{periodicmass}
M(x)=\left\{\begin{array}{cc}  +M,& jd \le x< (j+\frac12)d , \\
  -M,& (j+\frac12)d\le x<(j+1)d,
\end{array}\right.
\ee
where $d$ is the lattice period and $j\in\mathbb{Z}$ labels the unit cell.
For simplicity, we here assumed that the regions of positive and negative mass 
have the same spatial extent, $\ell\equiv d/2$, and the same absolute value 
of the mass, $|M(x)|=M$. This implies the symmetry $M(x+\ell)=-M(x)$.
Our calculations can easily be adapted to the general case, where 
we find that the spectrum acquires a gap, see Sec.~\ref{sec3c}.
For now, however, let us focus on Eq.~\eqref{periodicmass}.
In Sec.~\ref{sec3a}, we employ the transfer matrix method to solve 
the spectral problem and, in particular, to derive the energy quantization condition. 
The band structure and the corresponding Bloch states are described in Sec.~\ref{sec3b},
while we postpone the discussion of evanescent state solutions to Sec.~\ref{sec4}. 
Finally, in Sec.~\ref{sec3c}, a systematic low-energy theory is constructed by 
 projecting the model to the subspace spanned by the chiral zero modes. 

\subsection{Transfer matrix and spectral equation }\label{sec3a}

We first consider the unit cell $0<x<d$, where $\psi(d)$ and $\psi(0)$
are connected by the transfer matrix $T$,
\be\label{transferT}
\psi(d) = T \psi(0).
\ee
In this unit cell, Eq.~\eqref{W} implies that the wave function has the form 
\be\label{wf}
\psi(x) = \left\{\begin{array}{cc}W_M(x)  \begin{pmatrix}
a_1 \\ b_1\end{pmatrix}   &  \text{for} \; 0<x<\ell, \\
 W_{-M}(x) \begin{pmatrix}a_2 \\ b_2\end{pmatrix}   & \text{for} \; \ell < x < d,
\end{array}
\right. 
\ee
with $W_{\pm M}(x)$ in Eq.~\eqref{WM}.
The continuity of $\psi(x)$ at $x=\ell$ relates the complex coefficients $(a_2,b_2)$ and $(a_1,b_1)$ according to
\be\label{connection}
\begin{pmatrix} a_2 \\ b_2
\end{pmatrix} = W^{-1}_{-M}(\ell) \, W^{}_{M}(\ell) 
\begin{pmatrix} 
a_1 \\ b_1
\end{pmatrix}  ,
\ee
with  $W^{-1}_{-M}(\ell) \, W^{}_{M}(\ell)$ given in Eq.~\eqref{aux1}.
We can therefore express the transfer matrix as
\be\label{con2}
T = W_{-M}^{}(d) \, W_{-M}^{-1}(\ell) \, W_M^{}(\ell) \, W^{-1}_M(0).
\ee
The explicit form of the matrix elements of $T$ is given by Eq.~\eqref{Texplicit} in App.~\ref{appA}.
The matrix $T$ is symmetric and has $\det T=1$. Its eigenvalues 
can be written as $\lambda_\pm=e^{\pm iKd}$, 
where $K$ can be interpreted as a quasi-momentum along the $x$-direction.
As discussed below, $K$ can be either real-valued (for Bloch waves) or complex-valued (for evanescent modes).

In what follows, instead of $T$, we find it more convenient to use a 
modified transfer matrix $\Omega$ defined by
\be
T = W^{}_M(0) \, \Omega \, W^{-1}_M(0).
\ee
Using Eq.~\eqref{con2} and the relations $\psi(0)= W_{M}(0) \begin{pmatrix} a_1 \\ b_1
\end{pmatrix}$ and $\psi(d) = W_{-M}(d) \begin{pmatrix} 
a_2 \\ b_2 \end{pmatrix}$, which follow from Eq.~\eqref{wf},
we arrive at\footnote{With the matrix $D(x)={\rm diag}(e^{\kappa x},e^{-\kappa x})$ and
the matrix $\Omega_M$ for the single-kink problem in Eq.~\eqref{Om1}, we may express $\Omega$ as
$\Omega =  \Omega_M \, D(\ell) \, \Omega_M^{-1} \, D(\ell)$. This establishes a relation 
between the single-kink problem and the periodic problem.}
\be\label{omega}
\Omega =  W^{-1}_{M}(0) \, W_{-M}^{}(d) \, W^{-1}_{-M}(\ell) \, W_{M}^{}(\ell). 
\ee
The corresponding matrix elements are specified in Eq.~\eqref{Omegaexplicit}. 
We again have $\det \Omega=1$, and $\Omega$ has the same eigenvalues 
$\lambda_\pm=e^{\pm iKd}$ as $T$.  

We next require that $\psi(x)$ satisfies the Bloch periodicity condition
\be\label{Blochcondition}
\psi(x+d) = e^{i Kd} \,\psi(x), 
\ee
with a quasi-momentum $K$ along the $x$-direction. 
For Bloch wave solutions, $K$ must be real. We then take $K$ from the first BZ of the mass superlattice, 
\be\label{1BZ}
-\frac{\pi}{d} < K \leq \frac{\pi}{d},
\ee
where  $(K,k_y)=(0,0)$ is the ``$\Gamma$ point''.
More generally, we can impose Eq.~\eqref{Blochcondition} for complex values of $K$. 
We find three possible types of solutions, where $K$ is either real (Bloch waves)
or complex (evanescent waves), with $K=\pm i{\cal K}$ or $K=\mp i{\cal K}\pm \pi/d$.
The inverse length scale ${\cal K}>0$ is determined below.  
Evanescent state solutions thus are obtained
by imposing either
\be\label{imaginaryK}
\psi(x+d) = e^{ \mp \mathcal{K}d}\, \psi(x)
\ee
or 
\be\label{imaginaryKanti}
\psi(x+d) = - e^{ \pm \mathcal{K}d}\, \psi(x).
\ee
In what follows, evanescent waves derived from Eqs.~\eqref{imaginaryK} 
and \eqref{imaginaryKanti} are denoted as ``type-I'' and ``type-II'' states, respectively. 
While for the infinitely extended system evanescent states are not normalizable and hence not admissible, 
they emerge in the presence of boundaries or nonuniform potentials, see
Secs.~\ref{sec4} and \ref{sec5c}.

\begin{figure}[t!]
\centering
\includegraphics[width=\columnwidth]{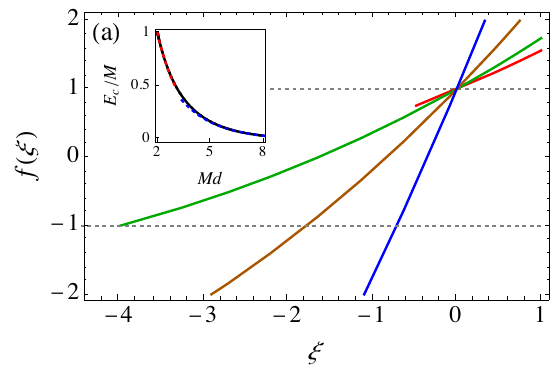}
\includegraphics[width=0.9\columnwidth]{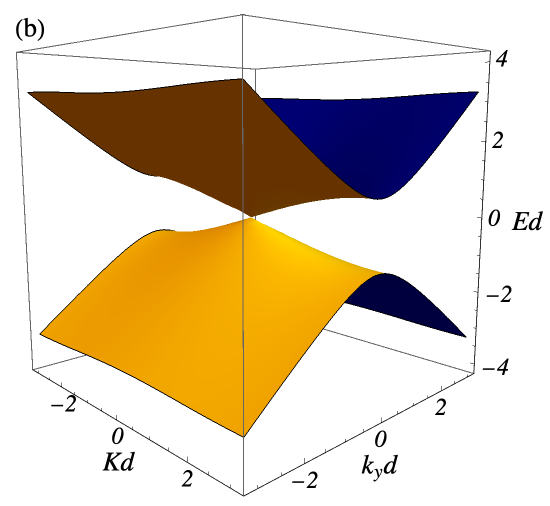}
\caption{Spectrum of the 2D Dirac Hamiltonian 
with the periodic mass term \eqref{periodicmass}.
(a) The function $f(\xi)$ vs $\xi$, see Eq.~\eqref{traceOmega}, in the regime $\xi>-(Md)^2$, 
for $Md=0.7, 2, 3.5$, and $5$, corresponding to the red, green, brown, and blue curves, respectively. 
According to Eq.~\eqref{tracecondition}, Bloch states require
$|f(\xi) |\le 1$.  For $f(\xi)>1$ [$f(\xi)<-1$], type-I [type-II] evanescent states are possible. 
Inset: Critical energy $E_c$ vs $Md$, where type-II states can only 
exist for $|E|>E_c$. The solid curve gives numerically exact results. 
The red and blue dotted curves give the analytical estimates
\eqref{Ec} for $Md \approx 2$ and $Md\gg 1$, respectively.
(b) Low-energy band structure, $E=\pm E_n(K,k_y)$, for Bloch states with $n=0$ and $Md=5$. }
\label{fig2}
\end{figure}

Setting $x=0$ and using the transfer matrix, Eq.~\eqref{Blochcondition} 
is next written as
\be
W_M^{}(0)\, \Omega \, W^{-1}_M(0)\,\psi(0) = e^{iKd}\psi(0),
\ee
which is equivalent to the condition
\be\label{wfcomp}
\left( \Omega -  e^{iKd} \mathbbm{1}\right) \begin{pmatrix} a_1\\
b_1
\end{pmatrix} = \begin{pmatrix} 0\\0\end{pmatrix}.
\ee
Nontrivial solutions of Eq.~\eqref{wfcomp} can only exist if
\begin{equation}\label{comp}
\det\left( \Omega -e^{i Kd }\mathbbm{1} \right) =0.
\end{equation}
The compatibility condition \eqref{comp} is equivalent to the
spectral equation
\be\label{tracecondition}
f(\xi) = \cos (Kd),
\ee
where we define $f(\xi) \equiv \frac12 \text{Tr}\, \Omega(\xi)$ with 
the dimensionless variable
\be\label{xidef}
\xi= (k_y^2-E^2)d^2.
\ee
Using Eq.~\eqref{Omegaexplicit}, one finds
\be\label{traceOmega}
f(\xi) =  \frac{(Md)^2+\xi\cosh\left(\sqrt{(Md)^2+\xi}\right)}{(Md)^2+\xi} .
\ee
The spectral equation thus depends on the single dimensionless parameter $Md$,
and $E$ and $k_y$ appear only through the dimensionless variable $\xi$.  
Below, 
we mostly focus on the low-energy regime, subject to the condition
\be\label{lowE}
|E|<M,
\ee
such that $\xi>-(Md)^2$. 
The function $f(\xi)$ is shown for several values of $Md$ in Fig.~\ref{fig2}(a).    
Bloch states are possible for $-1\le f(\xi)\le 1$ 
corresponding to $\xi_c\le \xi\le 0$, where $\xi_c<0$ is defined
by the condition $f(\xi_c)=-1$. 
Outside this window, no real solutions for the quasi-momentum $K$ can be found. 
However, Eq.~\eqref{tracecondition} also allows for solutions with complex-valued $K$.  
For $f(\xi)>1$, corresponding to $\xi>0$ and therefore $|E|<|k_y|$, 
we obtain type-I evanescent states.  On the other hand, for $\xi<\xi_c$,
we can have type-II evanescent states at energies above a critical
value, $|E|>E_c$ with $E_cd= \sqrt{-\xi_c}$, where we find the analytical estimate
\be \label{Ec}
    E_cd \approx \left\{ \begin{array}{cc}
    3-Md/2,  & Md \approx 2, \\
    2Md e^{-Md/2}, & Md \gg 1.
    \end{array}\right. 
\ee
In the low-energy regime \eqref{lowE}, solutions
for $\xi_c$, 
and thus type-II states, exist only for $Md>2$.  
This is related to the fact that 
if $Md<2$, for any Fermi level $|E_{\rm F}|<M$,
the Fermi surface is a closed curve in the 2D BZ. 
If $Md>2$, instead, the Fermi surface evolves from a closed curve (for $|E_{\rm F}|<E_c$)  
into a pair of disconnected arcs (for $E_c<|E_{\rm F}|<M$).
The critical point $|E_{\rm F}|=E_c$ corresponds to a Lifshitz transition.
Numerical results for $E_c$ vs $Md$ along with the estimates in Eq.~\eqref{Ec} 
are shown in the inset of Fig.~\ref{fig2}(a). 
For large $Md\gg 1$, type-II states are also realized at very low energies.

We discuss type-I and type-II states in more detail in Sec.~\ref{sec4} 
and focus on Bloch states with real $K$ for the remainder of this section.
We note in passing that Eq.~\eqref{tracecondition}  has also been specified 
in Ref.~\cite{zarenia2012}. However, 
the solutions $E=\pm \sqrt{k^2_y+M^2}$ reported in Ref.~\cite{zarenia2012}
are spurious, and the anisotropy of the emergent Dirac cone near
the $\Gamma$ point has been missed, see Eq.~\eqref{lowenergydisp} below.   
It is also worth mentioning that for $k_y=0$, Eq.~\eqref{tracecondition} coincides with the 
spectral equation for a generalized Kronig-Penney model 
of diatomic crystals \cite{Eldib1987,Smith2020}.

\subsection{Band structure and Bloch states}\label{sec3b}

We first study the solutions of the spectral condition \eqref{tracecondition} for real
quasi-momenta $K$ in the 1D BZ \eqref{1BZ}.  The corresponding Bloch bands form
the band structure of the mass superlattice.
For computing the band structure and the group velocities, it 
is convenient to introduce the auxiliary function
\be\label{phi}
\Phi(E,K,k_y) = f\left((k_y^2-E^2)d^2\right)-\cos(Kd),
\ee
where Eq.~\eqref{tracecondition} is equivalent to the condition $\Phi(E,K,k_y)=0$.
The band structure calculation amounts to finding the implicit function 
$E(K,k_y)$ defined by this condition.
In limiting cases, this can be done analytically (see below), but in general 
one has to resort to numerics.  In any case, one finds a 
particle-hole symmetric spectrum, $E=\pm E_n(K,k_y)$, where
 $n\in \mathbbm{Z}$ labels different bands with non-negative energy
$E_n(K,k_y)$.  
The group velocity $(v_x,v_y)$ for a 
given eigenstate follows with $E=\pm E_n(K,k_y)$ from Eq.~\eqref{phi} as  
\be
v_x = - \frac{\partial_K\Phi}{\partial_E\Phi},
\quad \label{groupvelocities}
v_y = - \frac{\partial_{k_y}\Phi}{\partial_E\Phi} .
\ee
The low-energy spectrum  determined numerically is shown in Fig.~\ref{fig2}(b).
To understand these results, we now examine limiting cases where analytical progress
is possible.

First, for $Md\to 0$, Eq.~\eqref{tracecondition} recovers
the standard isotropic massless Dirac cone with $k_x=K$ restricted to the first BZ \eqref{1BZ},
\be\label{masszero}
E= \pm E_{n}(K,k_y) = \pm \sqrt{ (K+ 2 \pi n/d)^2 + k_y^2},
\ee
which includes an isolated Dirac node at zero energy as well as finite-energy 
crossing points for $K=0$, because  $E_{n}(0,k_y)=E_{-n}(0,k_y)$,
and for $K=\frac{\pi}{d}$, because  $E_{n}(\frac{\pi}{d},k_y)=E_{-n-1}(\frac{\pi}{d},k_y)$. 
The finite-energy crossings points are not isolated but form lines when varying $k_y$.
We will show next that a finite value of $Md$ does not spoil the above
nodal structures at the center of the 1D BZ, but it does lift the degeneracies
at the BZ boundary where gaps open. 

For finite $Md$, let us first consider the 1D BZ center $K=0$.
We then find that Eq.~\eqref{tracecondition} has the non-negative solutions
\be \label{K=0}
E_{0}(0,k_y)= |k_y|, \quad
E_{n\ne 0}(0,k_y)= \sqrt{k_y^2  +\left(\frac{2\pi n}{d}\right)^2 + M^2}, 
\ee
where each energy $E_{n\ne 0}(0,k_y)$ is two-fold degenerate due to $\pm n$ bands. 
However, this degeneracy is lifted for $K\ne 0$, see Eq.~\eqref{localdisp} below.  
From Eq.~\eqref{K=0}, using $E^{(c)}_n\equiv E_{n}(0,0)$
for the $\Gamma$-point energy of the respective band,
 $\Gamma$-point crossings occur at zero energy ($n=0$) 
and at the finite energies $\pm E^{(c)}_{n\ne 0}$ with 
\be \label{crossingpoint}
E^{(c)}_{n\ne 0}=\sqrt{M^2+(2\pi n/d)^2}.
\ee

The zero-energy node is of special interest. 
By expanding Eq.~\eqref{tracecondition} for small energies and small momenta, 
one obtains an \emph{anisotropic} conical Dirac dispersion,
\be\label{lowenergydisp}
E=\pm E_{n= 0}(K,k_y) \simeq \pm \sqrt{ v^2_{x,0} K^2+v^2_{\rm F} k_y^2},
\ee
with a renormalized velocity along the $x$-direction,
\be\label{vx0}
\frac{v_{x,0}}{v_{\rm F}} =  \frac{Md/2}{\sinh(Md/2)}.
\ee
Numerical results for the full low-energy band structure are shown in Fig.~\ref{fig2}(b).
Near the $\Gamma$ point, they agree with Eq.~\eqref{lowenergydisp}.
Evidently, for $Md\to 0$, Eqs.~\eqref{lowenergydisp} and \eqref{vx0} recover the isotropic Dirac cone in Eq.~\eqref{masszero}.
For $Md\gg 1$, however, $v_{x,0}/v_{\rm F}$ is exponentially small 
and the dispersion becomes almost flat in the $K$-direction.  In this case, the 
individual mass kinks and anti-kinks in the periodic mass profile \eqref{periodicmass},
which are centered at $x=jd/2$ with integer $j$, bind 1D chiral zero modes by means of 
the Jackiw-Rebbi mechanism, see Sec.~\ref{sec2}.  
As we elaborate  in Sec.~\ref{sec3c},
superpositions of chiral zero modes generate the $n=0$ band dispersion \eqref{lowenergydisp}, 
where the finite hybridization between the counterpropagating zero modes at neighboring mass 
kinks and anti-kinks is responsible for the finite but exponentially small velocity \eqref{vx0}. 
While the anisotropic Dirac cone dispersion associated with zero modes in  
periodic mass profiles has been discussed before~\cite{Xiao2020}, the 
piece-wise constant mass term \eqref{periodicmass} admits an exact solution.  
We note that anisotropic Dirac cones can alternatively be engineered by means of scalar superlattice potentials
\cite{park2008,park2008b,Barbier2010,Li2017a,Li2021c} or by using periodic magnetic fields \cite{DellAnna2009,Tan2010,DellAnna2011}.

Similarly, we may expand around the $\Gamma$ point for the finite-energy crossing points 
\eqref{crossingpoint}, where we obtain 
\be \label{localdisp}
E_{n\ne 0}(K,k_y) \simeq E_n^{(c)} + \frac{k_y^2}{2E_n^{(c)}} + {\rm sgn}(n)\, v_{x,n}K,
\ee
with the velocities $v_{x,n\ne 0}=[2\pi n/(E^{(c)}_n d)]^2$ along the $x$-direction. 
We observe that a finite $k_y$ does not lift the two-fold degeneracy at $K=0$, and hence
there is a nodal line. 

Let us briefly compare the above results to the corresponding uniform-mass case $M(x)=M$, where
the band structure is given by
\be\label{BSconstantM}
E=\pm E^{(u)}_{n}(K,k_y)= \pm \sqrt{M^2 + (K+2\pi n/d)^2+ k_y^2 }.
\ee
Importantly, no zero-energy modes related to the anisotropic 
Dirac cone \eqref{lowenergydisp} appear anymore in Eq.~\eqref{BSconstantM}.  
Expanding around  the $\Gamma$ point,
where finite-energy crossings occur again at $E=\pm E_{n\ne 0}^{(c)}$ with $E_n^{(c)}$ in Eq.~\eqref{crossingpoint},
we find the positive-energy solutions
\bea\label{ppp}
E^{(u)}_{0}(K,k_y) &\simeq& M + \frac{k_y^2+K^2}{2M} , \\
E^{(u)}_{n\ne 0}(K,k_y) &\simeq& E_n^{(c)} + \frac{k_y^2}{2E_n^{(c)}}
+ \textrm{sgn}(n) \, \tilde v_{x,n} K, \nonumber
\eea
with $\tilde v_{x,n}=2\pi |n|/(E_n^{(c)} d)$.
The main difference between the alternating and the uniform mass profile is that 
the $n=0$ zero-mode band in Eq.~\eqref{lowenergydisp} has shifted to finite energies $E^{(u)}_0(K,k_y)\ge M$.
On the other hand,  the $n\ne 0$ dispersion relation \eqref{ppp} differs from 
Eq.~\eqref{localdisp} only with respect to the velocity  along the $x$-direction, $v_{x,n}\to \tilde v_{x,n}$. 

\begin{figure}[t!]
\centering
\includegraphics[width=.9\columnwidth]{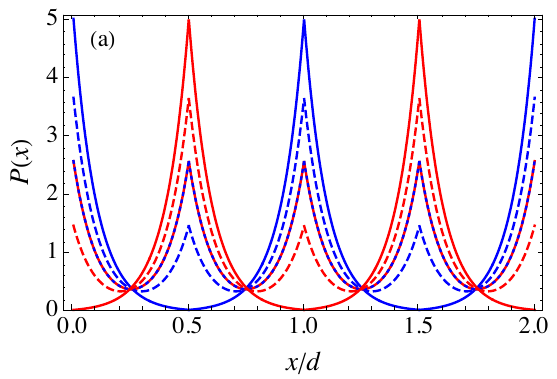}
\includegraphics[width=.95\columnwidth]{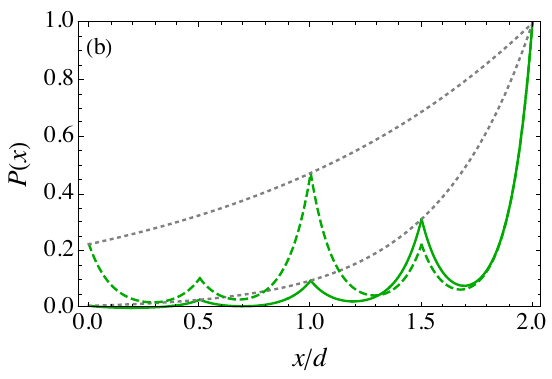}
\caption{Probability density $P(x)$ vs $x$ for selected eigenstates of the periodic mass problem.
(a) $P(x)$ for Bloch states with band index $n=0$, taking
$Md=5$ and $Ed=0.7$. Solid blue, dashed blue, blue-red, dashed red, and solid red curves
are for $k_yd=0.7, 0.3, 0, -0.3, -0.7$, respectively. 
(b) $P(x)$ normalized to its value at $x=2d$, for type-II evanescent states with 
$Md=5$ and $Ed=1$. Solid (dashed) green curves are for $k_yd=0$ ($k_y d=0.3$), while 
the dotted gray lines show the corresponding graphs of $e^{\mathcal{K}x}$.}
\label{fig3}
\end{figure}

Let us now turn to the Bloch eigenstates corresponding to the above band structure. 
Keeping $(E,k_y)$ implicit, we begin by expressing $\psi(x)$  in terms of a 
spinor wave function $u_K(x)$ with the periodicity of the mass superlattice,
\be
\psi(x) = e^{iKx}u_K(x),  \quad u_K(x+d)=u_K(x).
\ee
In the unit cell $0<x<d$, we obtain $u_K(x)=e^{-iKx}\psi(x)$ from $\psi(x)$ as specified
in Eq.~\eqref{wf}. We then need to determine the $K$-dependent 
coefficients $(a_1,b_1)$ and $(a_2,b_2)$ in Eq.~\eqref{wf}.  
To that end, we recall that $(a_2,b_2)$ follows from
$(a_1,b_1)$ by the continuity condition \eqref{connection} imposed at $x=d/2$.
Using Eq.~\eqref{wfcomp}, we can express\footnote{For  $K=0$, 
the matrix element $\Omega_{12}$ vanishes for the spectral branches $\pm E_0(0,k_y)=k_y$. 
Then Eq.~\eqref{b1a1} does not apply and we have instead 
$a_1=0$ with $b_1$ determined by normalization.}
$b_1$ in terms of $a_1$, 
\be\label{b1a1}
b_1(K) = \frac{e^{iKd}-\Omega_{11}}{\Omega_{12}} a_1,
\ee
with the  matrix elements of $\Omega$ in Eq.~\eqref{Omegaexplicit}.
Finally, $a_1$ is fixed by the normalization condition
\be
\label{norm}
\int_0^d dx\, |u_K(x)|^2 =1.
\ee
We thereby obtain the Bloch eigenstate $\Psi_{K,k_y,n,\pm}(x,y)=e^{i(Kx+k_y y)} 
u_{K,k_y,n,\pm}(x)$ for the energy $E=\pm E_n(K,k_y)$.
We illustrate the corresponding probability densities in Fig.~\ref{fig3}(a).
For $k_yd =0.7$ (solid blue curve), the state is mainly localized 
near the mass kinks at $x=jd$ with integer $j$.
For $k_yd =-0.7$ (solid red curve), on the other hand, the state is 
localized near the anti-kinks at $x=(j+1/2)d$.
As $|k_y d|$ decreases, one approaches the $d/2$-periodic probability density found for $k_y=0$, 
where the eigenstate is an equal-weight
superposition of counterpropagating chiral Jackiw-Rebbi modes.
 
For $E^2<k_y^2+M^2$ (where $\kappa$ is real),  we now observe that the particle current density \eqref{jx}
along the $x$-direction is uniform and given by  
\be\label{xcurrent}
j_x = \frac{-4\kappa \sin (Kd)}{(M+E)\Omega_{12}} |a_1|^2,
\ee
with $a_1$ determined by Eq.~\eqref{norm}.
Note that $j_x$ is odd in $K$. 
We note that
for the scattering problem in Sec.~\ref{sec5a}, instead of Eq.~\eqref{norm} it will be
more convenient to adopt a normalization where the wave function carries unit current. 
This is achieved by setting
\begin{equation}
  |a_1|^2=\left| \frac{(M+E)\Omega_{12}}{-4\kappa \sin (Kd)} \right|,
 \label{unitcurrentnorm}
\end{equation}
 which determines $a_1$, with $|a_1(-K)|^2=|a_1(K)|^2$,
 up to an irrelevant phase.

\subsection{Effective low-energy theory}\label{sec3c}

For $Md \gg 1$, the essential low-energy physics of the staggered
Dirac mass superlattice problem is captured by
projecting the full Hamiltonian \eqref{ham} onto the subspace spanned by
the 1D chiral zero modes centered at the (anti-)kink 
positions $x_j=jd/2$ (integer $j$) of the periodic mass term \eqref{periodicmass}. 
The resulting effective low-energy theory is also useful for
studying interacting variants of the model.
We show below that this projection reproduces the exact spectrum to 
exponential accuracy in the low-energy regime, $|E|< M$.

In the unit cell $|x|<\ell$ obtained after shifting $x\to x-\frac{\ell}{2}$, we start
from the mass profile 
\be\label{mass2}
M(x) = \left\{ \begin{array}{cc}
(1+\gamma) M,& |x|< \frac{\ell}{2}, \\
-(1-\gamma) M, &   \frac{\ell}{2} < |x|< \ell,
\end{array} \right. 
\ee
where $M>0$. The full mass profile follows by periodicity, $M(x+jd)=M(x)$, 
and is inversion symmetric, $M(x)=M(-x)$. We here allow for a dimensionless 
asymmetry parameter $\gamma$,
resulting in different mass amplitudes in regions of positive and negative mass.
Note that Eq.~\eqref{periodicmass} follows (up to the above shift) 
from Eq.~\eqref{mass2} for $\gamma=0$, where we also have 
$M(x+\ell)=-M(x)$.
The latter property is lost for $\gamma\ne 0$.
For $|\gamma|>1$, the mass term always has the same sign and chiral zero modes are absent. 
Below we focus on the more interesting case $|\gamma|<1$.

The kink and anti-kink positions in $M(x)$ define a 1D bipartite lattice in the $x$-direction, 
where sublattice $A$ (kinks) comprises the sites at $x_{Aj}=jd -\frac{\ell}{2}$ 
and sublattice $B$ (anti-kinks) refers to  $x_{Bj}=jd+\frac{\ell}{2}$. 
We now introduce the mass profile ${\cal M}_{\rm K}(x-x_A)$ for a
single kink centered at position $x_A$, and similarly ${\cal \bar M}_{\rm K}(x-x_B)$ for an anti-kink
centered at $x_B$, where 
\be\label{singlekink}
{\cal M}_{\rm K}(x) = M \textrm{sgn}(x) +\gamma M ,  \quad
{\cal \bar M}_{\rm K} (x)  ={\cal M}_{\rm K}(-x).
\ee \\
Zero-energy fermion modes bound to a kink or an anti-kink at $x=0$ satisfy
\bea \nonumber
\left(-i\sigma_x\partial_x  + {\cal M}_K(x) \sigma_z \right) \phi_+(x) &=&0, \\
\left(-i\sigma_x\partial_x  + \bar{\cal M}_K(x) \sigma_z \right) \phi_-(x) &=&0,
\eea
where the orthonormalized states $\phi_\pm(x)$ are eigenstates of $\sigma_y$
and satisfy $\phi_-(x) = \sigma_z \phi_+(-x)$.  Defining $\tilde M=(1-\gamma^2)M$, we find
\be\label{phipm}
\phi_\pm(x) = \sqrt{\frac{\tilde M}{2}}\, e^{-F(\pm x)} \begin{pmatrix}1 \\\pm i \end{pmatrix} ,
\quad F(x)=(|x|+\gamma x)M.
\ee
For constructing the low-energy theory for $Md\gg 1$, 
we expand the electron field operator in terms of the zero modes \eqref{phipm} 
for kink and anti-kinks centered at $x_{Aj}$ and $x_{Bj}$, respectively,
\be
\hat \Psi(x,y) = \sum_{j} [ \phi_+(x-x_{Aj})\,\hat \psi_{Aj}(y) + 
\phi_-(x-x_{Bj})\, \hat\psi_{Bj}(y) ] \label{fermionfield}
\ee
with 1D chiral fermion field operators $\hat\psi_{\alpha j}(y)$ 
for each sublattice $\alpha=A,B$ and each unit cell $j\in \mathbbm{Z}$ of the 1D bipartite lattice.
With fermion operators $c_{\alpha j k_y}$, we have
$\hat \psi_{\alpha j}(y)=\frac{1}{\sqrt{W}}\sum_{k_y} e^{ik_y y}\, c_{\alpha j k_y}$,
using periodic boundary conditions, $\hat\psi_{\alpha j}(y+W)=\hat\psi_{\alpha j}(y)$, such that 
$k_y=\frac{2\pi m}{W}$ for integer $m$ and linear system size $W$.

\begin{figure}[t!]
\includegraphics[width=0.9\columnwidth]{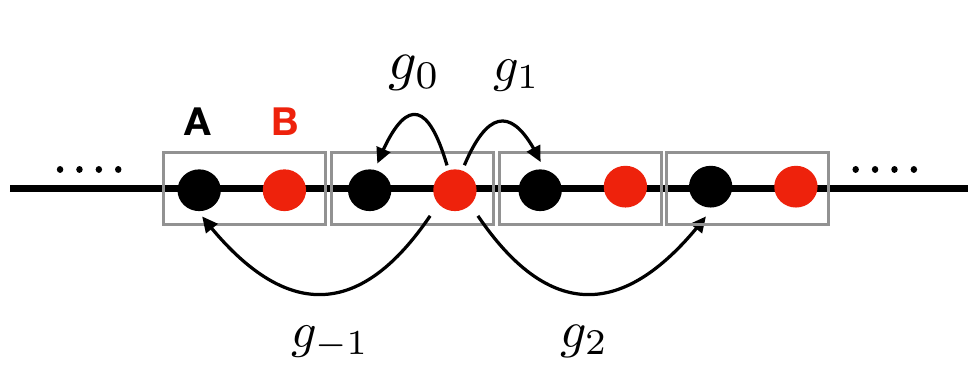} 
\caption{Illustration of the numbers $g_m$ in Eq.~\eqref{gm}, which encode the
overlap between counterpropagating chiral zero modes. The sites $A$ and $B$ correspond
to the 1D bipartite lattice of kink and anti-kink positions, where rectangles indicate
a unit cell. }
\label{fig4} 
\end{figure}

Projecting the full Hamiltonian $H$, see Eq.~\eqref{ham} with
$V(x)=0$ and $M(x)$ in Eq.~\eqref{mass2}, onto the low-energy basis \eqref{fermionfield},
we obtain the effective low-energy Hamiltonian,
\bea\label{projHam}
H_{\rm eff} &=& \int dx dy \,  \hat \Psi^\dagger(x,y)  H \hat \Psi(x,y) \\
&=& \nonumber \sum_{\alpha\alpha',jj',k_y}  
c^\dagger_{\alpha j k_y} {\cal H}^{\alpha\alpha'}_{jj'}(k_y)\, c^{}_{\alpha' j' k_y}, 
\eea
with the sublattice-diagonal matrix elements 
\bea\label{Haa}
&& {\cal H}^{AA}_{jj'}(k_y)  =
 k_y\tilde M \int dx \, e^{-F(x-x_{Aj}) -F(x-x_{Aj'})},\\ \nonumber
&&{\cal H}^{BB}_{jj'}(k_y) =
- k_y \tilde M \int dx \, e^{-F(-x+x_{Bj}) -F(-x+x_{Bj'})}.
\eea
Similarly, the off-diagonal components take the form
\bea
&& {\cal H}^{AB}_{jj'}(k_y)  = {\cal H}^{BA}_{j'j}(k_y)  =  \int dx \, e^{-F(x-x_{Aj})-F(-x+x_{Bj'})} \nonumber \\ 
&& \qquad \times\, \tilde M \left[  M(x) -\bar{\cal M}_{\rm K}(x-x_{Bj'}) \right].\label{Hab}
\eea
All matrix elements depend on the site indices $j$ and $j'$
only through their separation $(j-j')d$ and decay exponentially with this distance.
In particular, Eq.~\eqref{Haa} yields
\be\label{Haa2}
{\cal H}^{AA}_{jj'}(k_y)  = - {\cal H}^{BB}_{jj'}(k_y) = k_y \, f_{|j-j'|} ,  
\ee
where  the dimensionless numbers ($l=0,1,2,\ldots$)
\be\label{fldef}
f_l =\left(  \cosh(\gamma lMd ) +\frac{\sinh(\gamma l Md )}{\gamma} \right) e^{-lM d}
\ee
encode the overlap between zero-energy modes at distance $ld$ belonging to the same sublattice.
Note that $f_0=1$. The off-diagonal matrix elements \eqref{Hab} do not depend on $k_y$ and can similarly be
expressed as
\be\label{offd}
{\cal H}^{AB}_{jj'}(k_y) =M   g_{j-j'} ,  
\ee
where the dimensionless numbers $g_m$, with $m\in \mathbb{Z}$ 
and $M(x)$ in Eq.~\eqref{mass2}, are given by
\bea\nonumber
g_m  &= &\tilde M d\, e^{\gamma (m-\frac12)Md}\int ds \, e^{- (|s|+ |s-m+\frac12| ) Md} \\ 
&\times&  \left( \frac{ M(s+\frac14)}{M}+ \textrm{sgn}(s) -\gamma \right). \label{gm}
\eea
Note that for $\gamma=0$, we have $g_m=-g_{1-m}$.
For $Md\gg 1$, the numbers $f_l$ and $g_m$ decrease exponentially fast when 
increasing $l$ and $|m|$, respectively.  
The low-energy theory is dominated by terms with 
$f_{l=0}=1$ and $g_{m=0,1}$, corresponding to overlaps between at most adjacent sites of the 1D bipartite lattice, 
as illustrated in Fig.~\ref{fig4}.
In particular, the couplings $g_{-1}$ and $g_2$ describe next-nearest-neighbor overlap
integrals which are exponentially small compared to the nearest-neighbor couplings $g_{0,1}$, 
and can be omitted.
For $m=0,1$, the integral in Eq.~\eqref{gm} can be evaluated to exponential accuracy, 
\be \label{g01}
g_0 \approx -(1-\gamma^2) e^{-(1+\gamma)\frac{Md}{2}}, \quad
g_1 \approx (1-\gamma^2) e^{-(1-\gamma)\frac{Md}{2}}.
\ee

Since the matrix elements $\mathcal{H}^{\alpha\alpha'}_{jj'}(k_y)$ 
only depend on the separation $(j-j')d$, the low-energy Hamiltonian \eqref{projHam} 
is diagonal in momentum space.  Using the above chiral 1D fermion operators $c_{\alpha j k_y}$,
we define a momentum-space spinor field $C_{K k_y}$ according to
\be
\begin{pmatrix} c_{A j k_y} \\ c_{B j k_y} \end{pmatrix}=
 \int_{-\pi/d}^{\pi/d} \frac{dK}{2\pi} e^{ij Kd }  C_{ K k_y}, \quad
C_{Kk_y}=\begin{pmatrix} C_{AKk_y} \\ C_{BKk_y} \end{pmatrix}.
\ee
For  $W\to \infty$, we then obtain 
\be\label{efftheory}
H_{\rm eff} = \int \frac{dK dk_y}{(2\pi)^2}  \,
 C^\dagger_{Kk_y}\tilde {\cal H}(K,k_y)\, C_{Kk_y}^{},
\ee
where the single-particle effective Hamiltonian,
\be\label{Heff}
\tilde {\cal H}(K,k_y) = \begin{pmatrix} \tilde f(K) k_y & \tilde g(K)M\\
 \tilde g^*(K)M & - \tilde f(K) k_y 
 \end{pmatrix} ,
\ee
is expressed in terms of the Fourier series
\bea\label{fapp}
\tilde f(K) &=& f_0+2\sum_{l=1}^\infty f_l \cos(lKd) \approx 1, \\
\tilde g(K) &=& \sum_m g_m e^{-imKd}  \approx g_0+ g_1 e^{-iKd}. \nonumber 
\eea
The approximate results in Eq.~\eqref{fapp} are obtained by keeping only the 
leading coefficients $f_{l=0}=1$ and $g_{m=0,1}$,  
and hold to exponential accuracy for $Md \gg 1$.
By diagonalizing $\tilde {\cal H}(K,k_y)$ with the approximations in Eq.~\eqref{fapp},
we obtain the eigenenergies 
\be\label{zeromodeg}
    E(K,k_y) = \pm \sqrt{k^2_y + M^2[g_0^2+g_1^2 + 2g_0g_1 \cos(Kd)]} .
\ee
This expression accurately reproduces the $n=0$ band obtained 
from the exact spectral equation~\eqref{tracecondition}.

Close to the $\Gamma$-point ($Kd\ll 1$),  Eq.~\eqref{Heff} reduces to
\be \label{K=0ham}
  \tilde{\mathcal{H}} (K, k_y) =  Mg_1Kd\, \tau_y + k_y \tau_z -\left( \Delta +\frac12 Mg_1(Kd)^2\right)  \tau_x , 
\ee
with the gap $\Delta=-(g_0+g_1) M>0$ and Pauli matrices $\tau_a$ in sublattice space for the 1D bipartite lattice.  
Equation~\eqref{zeromodeg} then simplifies to the dispersion relation of anisotropic massive Dirac fermions,
\be
E(K,k_y)=  \pm \sqrt{\tilde v_x^2K^2+v^2_{\rm F}k_y^2+ \Delta^2 },\label{zeromodesp}
\ee 
with 
\be
\Delta = 2 \tilde M\, e^{-Md/2} \sinh(\gamma M d/2), \quad 
\frac{\tilde v_x}{v_{\rm F}} = \tilde M d \, e^{-M d/2}. \label{gap1}
\ee
For $\gamma=0$, we have $\Delta=0$ and Eq.~\eqref{zeromodesp} reproduces  Eq.~\eqref{lowenergydisp} since $v_{x,0}=\tilde v_x$ for $Md\gg 1$, see
Eq.~\eqref{vx0}. 
However, for $\gamma\neq 0$, the anisotropic Dirac cone is gapped 
and has the Chern number $C=-\frac12\,{\rm sgn}(\Delta)$ \cite{Alicea2012,Wang2021a,GirvinYang,Yao2009,Niu2010}.

We mention in passing that in terms of fermionic sublattice spinor fields,  
$\hat \psi_{j}(y)= \begin{pmatrix}\hat \psi_{Aj}(y)\\ \hat \psi_{Bj}(y)
\end{pmatrix},$ the low-energy Hamiltonian \eqref{efftheory} can also be written as 
\bea\nonumber
H_{\rm eff} &=& \sum_{j} \int dy  \Bigl \{ \hat \psi_j^\dagger \, [-i\partial_y \tau_z +Mg_0\tau_x ]\, \hat \psi_j^{} \\ &&\qquad + Mg_1 [\hat \psi_j^\dagger \tau_+ \hat\psi^{}_{j+1} + {\rm h.c.}]\Bigr\},
\eea
with the approximations in Eq.~\eqref{g01} and using $\tau_+=\frac12(\tau_x+i\tau_y)$. 
Such a representation can be useful in order to include, for instance, Coulomb interaction effects.

The above projection scheme can be adapted to any periodic mass profile $M(x)$ 
alternating between positive and negative values. 
For a continuous mass profile, the zeros of $M(x)$ define the sites of the 1D bipartite lattice, and
close to these zeros, a single (anti-)kink in Eq.~\eqref{singlekink} can be approximated by 
a linear function ${\cal M}_{\rm K}(x)= Mx/d$
($\bar{\cal  M}_{\rm K}(x) = - Mx/d$). In that case, the normalized zero-energy wave functions in Eq.~\eqref{phipm}
are replaced by
\be \label{genzeromodes}
\phi_\pm(x) = ( 4\pi M/d)^{-1/4}\, e^{-\frac{M}{2d} x^2} \begin{pmatrix}1 \\\pm i \end{pmatrix}.
\ee
The effective low-energy Hamiltonian is then still given by Eq.~\eqref{Heff}, with 
 $\tilde f(K)$ and $\tilde g(K)$ now calculated with $\phi_\pm$ in Eq.~\eqref{genzeromodes}.
We conclude that the projection approach offers a powerful route towards studying the
low-energy theory of Dirac fermions in a mass superlattice. 

\section{Boundary modes}
\label{sec4}

We now turn to evanescent wave solutions which are characterized by a 
complex-valued quasi-momentum $K$ and can arise in the presence
of boundaries or nonuniform potentials.
Throughout this section, we focus on boundary-induced evanescent states in a constant potential
and set $V(x)=0$.  In addition, we consider the low-energy regime \eqref{lowE}, 
where $\kappa$ in Eq.~\eqref{kappadef} is real-valued
and $(Md)^2+\xi>0$ in Eq.~\eqref{traceOmega}. 
The length scale $\kappa^{-1}$ governs the decay (or growth) 
of the wave function along the $x$-direction in a region of constant mass. 
For the piece-wise constant mass term \eqref{periodicmass}, 
the length $\kappa^{-1}$ thus represents a microscopic scale, 
which is only relevant on scales below the period $d$ and which becomes shorter with increasing $|k_y|$.
As discussed below, the mass superlattice  generates another 
characteristic length scale, ${\cal K}^{-1}$, which governs the decrease (or increase) 
of evanescent waves on scales larger than the superlattice period $d$ and which, for small $|k_y|$,
grows with increasing $|k_y|$.  
In Sec.~\ref{sec4a}, we summarize general properties of evanescent states,
followed by the explicit calculation of boundary modes for a semi-infinite geometry in Sec.~\ref{sec4b}.

\subsection{Evanescent states}\label{sec4a}

\begin{figure}[t]
\includegraphics[width=\columnwidth]{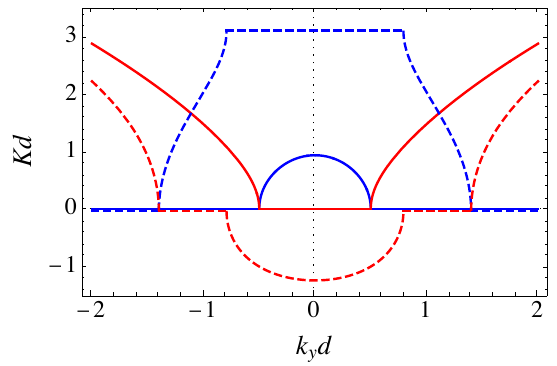}
\caption{Quasi-momentum $K$ in Eq.~\eqref{K} vs $k_y$ for $Md=4$, taking 
the $+$ sign in Eq.~\eqref{K}. Red (blue) curves show the imaginary (real) part of $K$.
The solid curves are for $Ed= 0.5$ and the dashed curves for $Ed=1.4$. 
}
\label{fig5}
\end{figure}

The spectral condition \eqref{tracecondition} is formally solved by 
\be\label{K}
Kd=\pm \arccos f(\xi) ,
\ee
with the function $f(\xi)$ in Eq.~\eqref{traceOmega}.  
Bloch wave solutions with real $K$ only exist for $|f(\xi)| \le 1$. 
For $f(\xi)>1$, corresponding to $\xi>0$ and thus to $|E|<|k_y|$, 
one instead finds a purely imaginary solution, $K=\pm i{\cal K}$ 
for the respective sign in Eq.~\eqref{K}, with the convention ${\cal K}>0$.   
For $0<\xi \ll 1$, we estimate
\be \label{KtypeI}
  \mathcal{K}d \simeq \frac{\sinh(Md/2)}{Md/2} \sqrt{\xi} ,
\ee
in agreement with Eqs.~\eqref{lowenergydisp} and \eqref{vx0}.  
The resulting type-I boundary modes, see Eq.~\eqref{imaginaryK}, originate from states near the superlattice BZ 
center and are directly connected to the anisotropic Dirac cone dispersion \eqref{lowenergydisp}.  
This case is illustrated for $Ed=0.5$ (solid curves) in Fig.~\ref{fig5}.   
For the wave function \eqref{wf} of type-I states, using Eqs.~\eqref{wfcomp} and \eqref{Omegaexplicit}, we obtain 
\be\label{b1a1a}
\frac{b_1(K=\pm i\mathcal{K})}{a_1}= \frac{e^{\mp\mathcal{K}d}-\Omega_{11}}{\Omega_{12}},
\ee
resulting in a decay (increase) of $\psi_K(x)$ with increasing $x$ for $K=i\mathcal{K}$ ($K=-i{\cal K}$).
We note that the particle current along the $x$-direction vanishes, $j_x=0$, because $b_1/a_1$ is real.

Next we turn to the case $Ed=1.4$ (dashed curves in Fig.~\ref{fig5}), where 
the real part of $K$ again vanishes for $|E|<|k_y|$, corresponding to type-I states. 
However,  for small $|k_y|$ and $Md>2$, a region with $f(\xi)<-1$ corresponding to
$\xi<\xi_c<0$ exists, cf.~Fig.~\ref{fig2}(a), where Eq.~\eqref{K} yields a pair of type-II states 
 with $K=\mp i\mathcal{K}\pm \pi/d$, see Eq.~\eqref{imaginaryKanti}. For $\xi \lesssim\xi_c$, we find
\begin{equation}\label{KtypeII}
\mathcal{K}d \simeq \sqrt{f'(\xi_c)\, (\xi_c-\xi)}.
\end{equation}
From Eq.~\eqref{KtypeII} and Fig.~\ref{fig5}, we observe that the decay length ${\cal K}^{-1}$ can 
exceed the lattice spacing $d$ of the mass superlattice.
The wave function of type-II states also follows from Eq.~\eqref{wf} but with
\be\label{b1a1b}
\frac{b_1\left(K=\mp i\mathcal{K}\pm \pi/d\right) }{a_1}= \frac{-e^{\pm\mathcal{K}d}-\Omega_{11}}{\Omega_{12}},
\ee
again resulting in $j_x=0$.  The corresponding spatial probability density is
illustrated in Fig.~\ref{fig3}(b), where an overall decay on the emergent (long) length scale ${\cal K}^{-1}$ 
is clearly visible. At the same time, the microscopic length $\ell=d/2$ due to the mass superlattice causes
a periodic modulation of the spatial decay.

The emergence of type-II states can also be seen from the results
of Sec.~\ref{sec3c}.  Near the boundary of the superlattice BZ, by 
writing $K =\frac{\pi}{d} + q$ with $|q|d\ll 1$,
the low-energy dispersion relation \eqref{zeromodeg} takes the form
\be\label{EboundaryBZ} 
E\left(\frac{\pi}{d}+q ,k_y\right) \approx \pm \sqrt{- \tilde v_x^2 q^2 + k^2_y + E_c^2},
\ee
with $\tilde v_x$ in Eq.~\eqref{gap1} and $E_c= 2\tilde M e^{-\frac{Md}{2}}\cosh(\frac{\gamma Md}{2})$.
Equation~\eqref{EboundaryBZ} reveals a saddle point at the BZ boundary, which is responsible for 
the Lifshitz transition discussed in Sec.~\ref{sec2a}.
For $|E|<E_c$, Bloch states with real $q$ exist for any (small) value of $k_y$. However, for 
$|E|>E_c$, type-II states with imaginary $q$ emerge for $k_y^2< E^2-E_c^2$.

\subsection{Boundary modes for semi-infinite geometry}\label{sec4b}

It is instructive to study a specific example admitting evanescent wave solutions. 
We here consider  the Dirac mass superlattice problem on the half-plane $x<x_0$, 
with the boundary line $x=x_0$ located in a positive-mass region, 
say, $0<x_0<\frac{d}{2}$. 
We impose a boundary condition at $x=x_0$ and $y\in \mathbb{R}$,
\be
\mathcal{B}(\alpha)\, \Psi(x_0,y) = \pm \Psi(x_0,y),
\label{bc}
\ee
 which ensures that the component of the current density normal to the boundary vanishes \cite{Witten2016,buccheri2022}.
The matrix ${\cal B}$ depends on a phenomenological boundary angle $\alpha$,
\be
\mathcal{B}(\alpha) = \sigma_y \cos \alpha + \sigma_z \sin \alpha.
\ee
For definiteness, we choose the eigenvalue $+1$ in Eq.~\eqref{bc} in what follows. (The solution 
for eigenvalue $-1$ follows by replacing $\alpha \to \alpha+\pi$.)
The corresponding eigenstate of $\mathcal{B}(\alpha)$ is given by
\be
|\alpha\rangle = \begin{pmatrix} \cos(\frac{\alpha}{2}-\frac{\pi}{4}) \\ -i \sin(\frac{\alpha}{2}-\frac{\pi}{4}) \end{pmatrix}.
\label{Beigen}
\ee
We now consider parameter regions with $|f(\xi)| >1$, where Bloch waves are absent and $K$ in Eq.~\eqref{K} is complex-valued.  

\begin{figure}[t]
\includegraphics[width=\columnwidth]{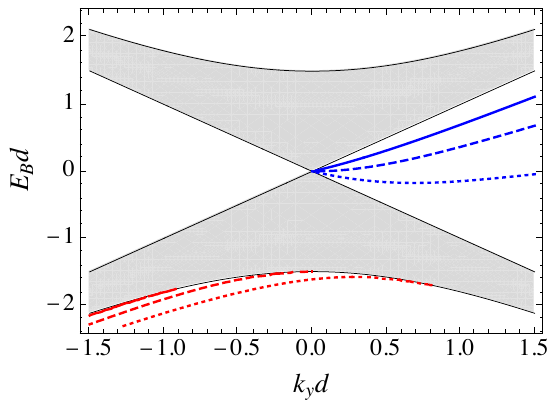}
\caption{Dispersion relation $E_B(k_y)$ of type-I (blue) and type-II (red) boundary modes
in a semi-infinite geometry with $x<x_0$.  We assume $x_0=d/4$ and $Md=3.1$, where 
results obtained by numerically solving Eq.~\eqref{boundcond}
are shown for the boundary angles $\alpha=\pi/3, \pi/2,$ and $2\pi/3$, using solid, dashed, and 
dotted lines, respectively.  The shaded region corresponds to Bloch states. }
\label{fig6}
\end{figure}

For the semi-infinite problem, 
normalizable states can be obtained only from one of the two solutions in Eq.~\eqref{K}.  
Denoting this solution by $K=K_0$ and recalling our convention ${\cal K}>0$, we have
$K_0=-i{\cal K}$ for type-I states with $\xi>0$.
Similarly, we have $K_0=-i{\cal K}+\pi/d$ 
for type-II states with $\xi<\xi_c<0$.  
For $x\to -\infty$, the solution $\psi_{K_0}(x)$ decreases exponentially and 
therefore  describes a normalizable state. 
The other solution $\psi_{-K_0}(x)$  grows exponentially for $x\to -\infty$ and 
hence is not admissible. 

The boundary condition \eqref{bc} implies that the boundary spinor $\psi_{K_0}(x_0)$ 
must be proportional to the state $|\alpha\rangle$ in Eq.~\eqref{Beigen}.
Using Eq.~\eqref{wfcomp}, we thereby arrive at the spectral condition
\be\label{boundcond}
\left( \Omega- e^{iK_0 d} \mathbbm{1} \right) W_M^{-1}(x_0) |\alpha \rangle =0,
\ee
which determines the dispersion relation of the boundary modes $E=E_B(k_y)$.
We illustrate typical results in Fig.~\ref{fig6}
for different values of the boundary angle. For $Md>2$, 
we observe both type-I boundary modes with $|E_B(k_y)|<|k_y|$ 
(blue curves) and type-II boundary modes (red curves). 
In both cases, the precise shape of the dispersion $E_B(k_y)$ sensitively 
depends on the angle $\alpha$ and on the boundary location $x_0$ (not shown). Moreover, 
the dispersion is not symmetric in $k_y$, which implies that  
the boundary modes can carry unidirectional currents.
We therefore expect them to  be observable in transport experiments.  
In addition, they could be detected in STM experiments.

\section{Potential step and interface modes}\label{sec5}

In this section, we return to the extended problem (without boundaries) for the Dirac Hamiltonian~\eqref{ham} 
with the periodic mass term in Eq.~\eqref{periodicmass}.  
We now include an electrostatic potential step of moderate step size $2V_s$ at position $x=x_s$,
\begin{equation}\label{step}
V(x)=V_s \; {\rm sgn}(x-x_s),\quad 0<2V_s<M.
\end{equation}
The potential \eqref{step} defines an $np$-junction. For definiteness, 
we assume $0<x_s<\frac{d}{2}$ such that the step is located in a region 
of positive mass. 

Here we focus on the most interesting low-energy regime with 
real-valued $\kappa$ parameters in Eq.~\eqref{kappadef}. 
Recalling that a uniform potential can be accounted for by shifting the energy $E$,
on the left side $x<x_s$, $\kappa=\kappa_L$ follows from Eq.~\eqref{kappadef} with $E\to E+V_s$ .
Similarly, $\kappa=\kappa_R$ for $x>x_s$ is obtained by replacing $E\to E-V_s$. 
(Below we will also use $\xi_{L,R}$ which follows from Eq.~\eqref{xidef} with the same
substitutions.)
In order to have both $\kappa_L$ and $\kappa_R$ real for all values of $k_y$, we require
\begin{equation}\label{zeromodesc}
    |E|<M-V_s.
\end{equation}
Apart from evanescent states bound to the potential step, we then have to take into account 
only the zero-mode band with $n=0$ corresponding to the emergent
anisotropic Dirac cone near the $\Gamma$ point. 

In Sec.~\ref{sec5a}, we consider scattering states and 
calculate the corresponding transmission probability for the potential step \eqref{step}.
The linear two-terminal conductance $G$ is discussed in Sec.~\ref{sec5b}, where we 
consider transport across the junction with lead electrodes attached to the system at $x\to \pm \infty$.  
Interestingly, we find a pronounced dependence of $G$ on the step position $x_s$. 
In Sec.~\ref{sec5c}, we then determine the dispersion relation of interface modes, which are
spatially localized near the potential step in the $x$-direction but propagate along the $y$-direction. 

\subsection{Scattering states and transmission probability}\label{sec5a}

We here consider scattering states with energy 
\begin{equation}
\label{phscattering}
|E| <V_s.
\end{equation}
Since the emergent Dirac cones on the two sides of the junction 
are shifted by the potential in opposite directions, in this energy window
one finds a particle-like state on the left side  and a hole-like state on 
the right side of the $np$-junction. The associated group velocity is then 
parallel (anti-parallel) to the momentum $K$ on the left (right) side. We note that for $0<2V_s<M$,  
Eq.~\eqref{phscattering} automatically implies Eq.~\eqref{zeromodesc}.  
For given $E$ and $k_y$, we have a pair of 1D Fermi momenta $\pm K_L$  on the left side, 
and similarly  $\pm K_R$ on the right side. The values of $K_L>0$ and $K_R>0$
follow from the spectral equation \eqref{tracecondition}. In particular, using the auxiliary function
$\Phi(E,K,k_y)$ in Eq.~\eqref{phi}, $K_{L,R}$ are the solutions of
\be\label{defKL}
\Phi(E+V_s,K_L,k_y) =0, \quad \Phi(E-V_s,K_R,k_y) =0.
\ee
We then use Eqs.~\eqref{wf} and \eqref{b1a1} to determine the scattering state 
by matching the wave function on the left side of the junction 
to the wave function on the right side.
Appending energy arguments as indices on the matrix $W_M(x)$ in Eq.~\eqref{WM},
the full wave function for $0<x<\frac{d}{2}$ is written as 
\bea\nonumber
\psi(x<x_s) &=& W_{E+V_s,M}(x) \left[ \begin{pmatrix}
a_1 \\ b_1
\end{pmatrix}_{K_L} + r \begin{pmatrix}
a_1 \\ b_1
\end{pmatrix}_{-K_L}  \right], \\
\label{wfL} 
\psi(x>x_s) &=& t\, W_{E-V_s,M}(x) 
\begin{pmatrix}
a_1 \\ b_1
\end{pmatrix}_{-K_R},
\eea
with complex-valued reflection $(r$) and transmission ($t$) amplitudes. 
We normalize the incident, reflected, and transmitted wave functions
such that they carry unit current, see Eq.~\eqref{unitcurrentnorm}.
Notice that the wave function for $x>x_s$ describes a hole propagating to the right and therefore involves the 1D Fermi momentum $-K_R$.

The transmission probability $\mathcal{T}$ is given by
\begin{equation}\label{transmissionprob}
\mathcal{T}(E,k_y)= |t|^2= \left| \frac{a_1(K_L)}{a_1(K_R)} \right|^2 |t'|^2,
\end{equation}
where the amplitude $t'$ follows by setting all 
coefficients $a_{1}(\pm K_{L,R})=1$ in Eq.~\eqref{wfL}.  
Continuity of $\psi(x)$  at $x=x_s$ then results in 
two coupled linear equations for  $r$ and $t'$,
\bea \nonumber
&&
W_{E+V_s,M}(x_s) \left[ \begin{pmatrix}
1 \\ b_1
\end{pmatrix}_{K_L} + r \begin{pmatrix}
1 \\ b_1
\end{pmatrix}_{-K_L}  \right]=\\
&& = t'\, W_{E-V_s,M}(x_s) 
\begin{pmatrix}
1 \\ b_1
\end{pmatrix}_{-K_R},
\label{contcond}
\eea
where, using $a_1=1$, Eq.~\eqref{b1a1} gives 
\be\label{b1K}
b_1(\pm K_L)=\frac{e^{\pm iK_L d}-\Omega_{11}(E+V_s)}{\Omega_{12}(E+V_s)},
\ee
and analogously for $b_1(\pm K_R)$. 
Note that the energy argument of the $\Omega$ matrix elements \eqref{Omegaexplicit} 
has been made explicit.  With the auxiliary quantities 
\be\label{AB}
\begin{pmatrix}
A(K) \\B(K)
\end{pmatrix}= W^{-1}_{E+V_s,M}(x_s)\, W_{E-V_s,M}^{}(x_s) 
\begin{pmatrix}
1 \\ b_1(K)
\end{pmatrix},
\ee
where we suppress the dependence on $x_s$,
reflection and transmission amplitudes can be expressed as
\bea\label{reflampl}
r&=& - \frac{B(-K_R)-b_1(K_L)A(-K_R)}{B(-K_R)-b_1(-K_L)A(-K_R)},\\ \nonumber
t'& =& \frac{b_1(K_L)-b_1(-K_L)}{B(-K_R)-b_1(-K_L)A(-K_R)}. 
\eea
We thus obtain the reflection probability ${\cal R}=|r|^2$ and the
transmission probability ${\cal T}$ from Eq.~\eqref{transmissionprob}.
Of course, current conservation yields ${\cal T}=1-{\cal R}$.  

\begin{figure}
\includegraphics[width=0.95\columnwidth]{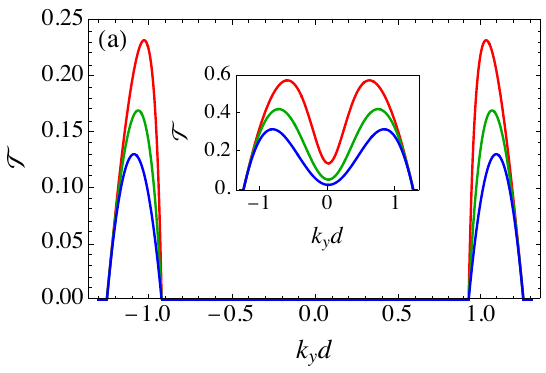}
\includegraphics[width=0.95\columnwidth]{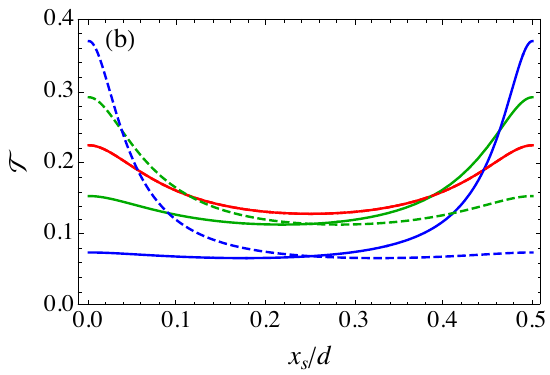}
\caption{
Transmission probability ${\cal T}$ for the Dirac mass superlattice 
in the presence of the potential step \eqref{step} with $V_sd=1.25$. 
(a) $\mathcal{T}$ vs $k_y$ for $E=0$, $Md=5$, 
and $x_s/d= 0.05, 0.1, 0.25$ (red, green, blue curves). Inset: Same parameters as in the main panel 
but for $Md=3.7$. (b) $\mathcal{T}$ vs step position $x_s$ for $Md=5$ with $k_yd=1.1$ (solid lines) 
and $k_yd=-1.1$ (dashed lines), using $Ed=0, 0.05, 0.1$ (red, green, blue curves).  
}
\label{fig7}
\end{figure}

We illustrate typical results for the transmission probability in Fig.~\ref{fig7}. 
Depending on the parameters, Bloch states, and thus a finite transmission,  
can only be realized in a window of $k_y$ values.
For fixed step position, we indeed observe a strong dependence on $k_y$, 
with the symmetry $\mathcal{T}(E=0,-k_y)  =\mathcal{T}(E=0,k_y)$, 
cf.~Fig.~\ref{fig7}(a), 
where we also illustrate the effect of changing the parameter $Md$. 
In particular, we see that at fixed energy, for the case of larger mass 
in the main panel of Fig.~\ref{fig7}(a), 
there is a window around $k_y=0$ where the transmission vanishes. 
This window shrinks as the mass decreases, and eventually closes, as shown in the inset. 
Notice that the window's edges do not depend on the position of the step. For fixed $(E,k_y)$, 
Fig.~\ref{fig7}(b) reveals a pronounced dependence of ${\cal T}$ on the step position $x_s$, 
with the symmetry $\mathcal{T}(\frac{d}{2}-x_s,k_y)=\mathcal{T}(x_s,-k_y)$.
This effect is linked to the strong $x$-dependence of the wave functions. Indeed, as discussed
in Sec.~\ref{sec3c}, the low-energy  states are built from chiral 
zero modes which are localized along the $x$-direction near $x=jd/2$ (integer $j$).  
Depending on the sign of $k_y$, we find high transmission probability if $x_s$ 
is near one of these positions, where the probability density has maxima, see Fig.~\ref{fig3}(a).
In the next section, we study how this behavior affects the electrical conductance. 

\begin{figure}
\includegraphics[width=\columnwidth]{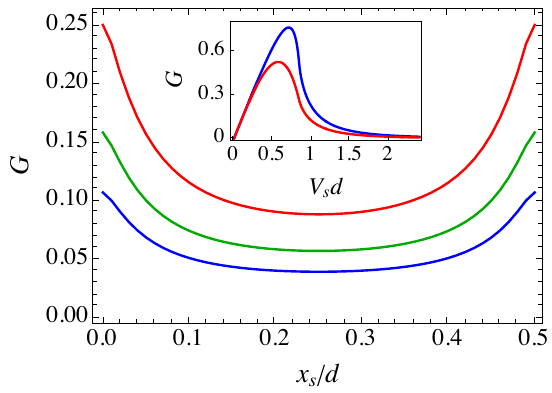}
\caption{Conductance $G$ for the Dirac mass superlattice with $Md=5$ at Fermi energy $E_{\rm F}=0$
in the presence of the potential step \eqref{step}. 
We show $G$ in units of $\frac{N_v e^2 W}{(2\pi)^2 \hbar d}$ for a strip of width $W$
and degeneracy index $N_v.$ Main panel: $G$ vs step position $x_s$ for several values of the 
potential step size, $V_sd=1.1, 1.25, 1.4$, shown by red, green, and blue curves, respectively.
Inset: $G$ vs $V_s$ for $x_s=0.05d$ (blue) and $x_s=0.25d$ (red curve).}
\label{fig8} 
\end{figure}

\subsection{Conductance}
\label{sec5b}

Within a noninteracting theory, the transmission probability ${\cal T}(E,k_y)$ 
directly determines the linear two-terminal conductance $G$ via the standard
Landauer-B\"uttiker formula \cite{datta_1995}. At zero temperature,  
identifying $E$ with the Fermi energy $E_{\rm F}$, the conductance for a 
strip of large width $W$ along the $y$-direction, with source and drain electrodes
adiabatically connected at $x\to \pm \infty$, is given by
\begin{equation}\label{conductance}
G = \frac{N_ve^2 W}{(2\pi)^2 \hbar} \int dk_y  \, \mathcal{T} (E_{\rm F},k_y),
\end{equation}
where $N_v$ is a degeneracy factor. For instance, in a graphene monolayer, we have
$N_v=4$ because of spin and valley degeneracies. Note that at given energy,
only states with $k_y$  such that $|f(\xi_{L,R})|< 1$
have finite transmission probability and contribute to the conductance. 

\begin{figure*}
\centering
\includegraphics[width=4.3cm]{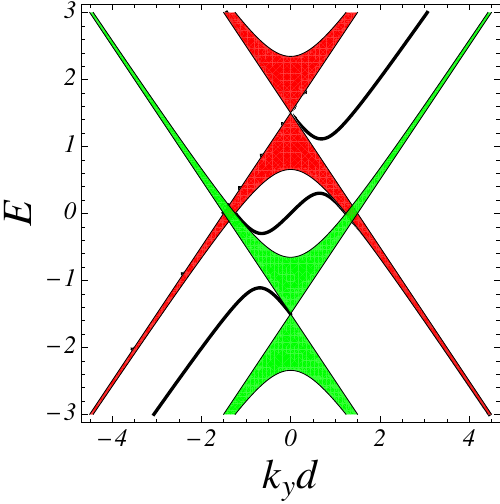}~~~\includegraphics[width=4.3cm]{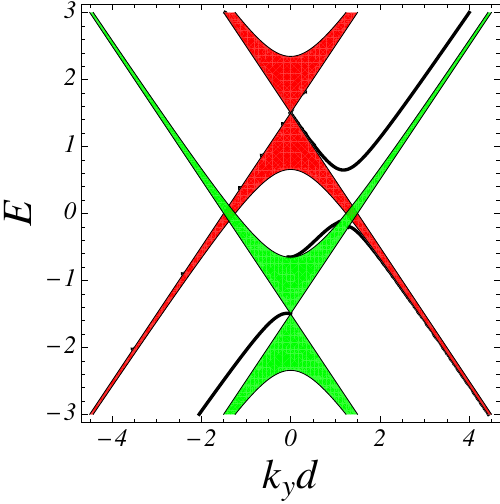}~~~\includegraphics[width=4.3cm]{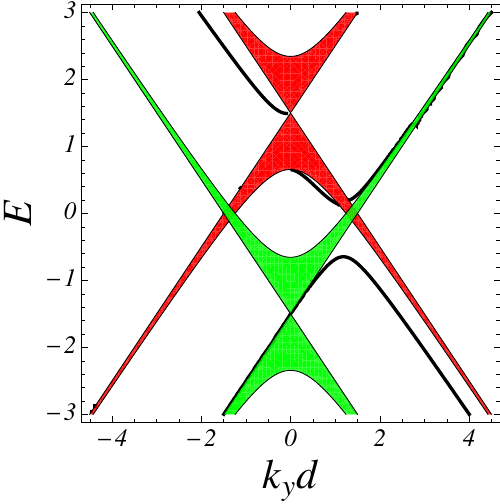}~~~\includegraphics[width=4.3cm]{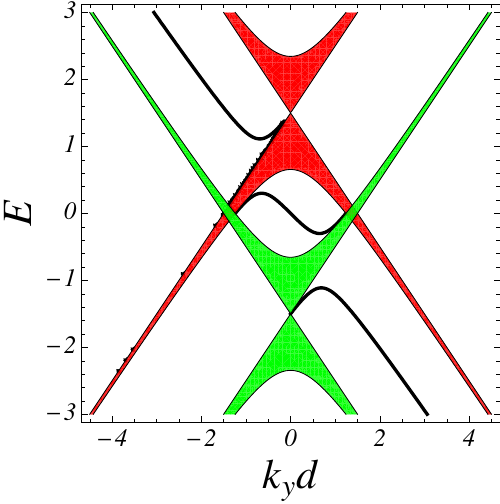}
\caption{Dispersion relation of interface states bound to a potential step  with $V_sd=1.5$ and several
step positions $x_s$ for $Md=5$, with $E$ in units of $\hbar v_{\rm F}/d$. 
The green (red) bands correspond to Bloch states at $x<x_s$ ($x>x_s$). 
The solid black curves refer to interface modes.  The interface modes in the
central inner region are of type II-II, while all others are of type I-II. 
From left to right panel: $x_s=0$, $x_s=0.1d$, $x_s=0.4d$, and $x_s=0.5d$.}
\label{fig9}
\end{figure*}

We illustrate the dependence of $G$ on the potential step position $x_s$ 
and on the step size $V_s$ in Fig.~\ref{fig8}. We observe that $G$ strongly depends on $x_s$ and, 
in the interval $0<x_s<d/2$, exhibits a broad minimum at $x_s=d/4$
with the symmetry $G(\frac{d}{2}-x_s)=G(x_s)$.
The conductance will then be a periodic function of $x_s$ with period $d/2$. 
Such conductance oscillations are most pronounced for $Md\gg 1$ and small values of the Fermi energy,
where the relevant electronic states originate from the chiral zero modes 
localized near the mass (anti-)kinks at $x=j d/2$. The $x_s$-dependence of $G$ becomes weaker for smaller values of $M$ (results not shown).
A pronounced spatial dependence of $G$ on the step position is therefore 
an hallmark of the existence of zero modes which are well localized along the $x$-direction.

As a function of  step size $V_s$, 
we observe that the conductance shows a broad peak, cf.~inset of Fig.~\ref{fig8}. 
This behavior can be rationalized by noting that in this example
we consider $E_{\rm F}=0$, where the density of states associated with the Dirac cone, 
and hence also the conductance, vanishes for $V_s\to 0$. Moreover, upon increasing $V_s$, 
the phase space for transmission (the window of $k_y$ where the transmission amplitude is
finite) first increases, but eventually shrinks and, as a consequence, the conductance 
decreases toward zero.

\subsection{Interface states}
\label{sec5c}

We finally study states localized near the interface at $x=x_s$.
These states are formed by a combination of either type-I or type-II evanescent waves 
on opposite sides of the step, matched at $x=x_s$.
In particular, solutions with type-II modes on both sides 
(``type II-II'' interface modes) require $f(\xi_{L})<-1$ and $f(\xi_R)<-1$, and 
have quasimomenta
\be \label{KLR}
K_{L}=- i{\cal K}_{L}+\frac{\pi}{d},\quad K_R=+i{\cal K}_R-\frac{\pi}{d},
\ee
with ${\cal K}_{L,R}>0$ given by Eq.~\eqref{KtypeII} with the replacement $E\to E\pm V_s$. 
The state $\psi_{K_L}(x)$ (for $x<x_s$) then shows
an exponential decay for $x\to -\infty$ and, 
similarly, $\psi_{K_R}(x)$ (for $x>x_s$) decays for $x\to \infty$. 
For type I-II interface states, composed of type-I and type-II modes on opposite 
sides, we find that, for $E>0$, the type-II state is on the left 
and the type-I on the right, with
\be 
K_{L}=- i{\cal K}_{L}+\frac{\pi}{d},\quad K_R=+i{\cal K}_R,
\ee
while for $E<0$, the opposite happens, with 
\be 
K_{L}=- i{\cal K}_{L},\quad K_R=+i{\cal K}_R -\frac{\pi}{d}.
\ee
The wave function matching condition at $x=x_s$ now implies 
\be
W_{E+V_s,M} (x_s) \begin{pmatrix}
a_1 \\ b_1
\end{pmatrix}_{K_L} = t'\, W_{E-V_s,M} (x_s) \begin{pmatrix}
a_1 \\ b_1
\end{pmatrix}_{K_R},
\ee
with $b_1(K)$ in Eq.~\eqref{b1K}.  
Using the auxiliary quantities in Eq.~\eqref{AB}, 
we arrive at the equation
\be\label{interfacecond}
\left( \Omega(E+V_s) - e^{iK_Ld} \mathbbm{1} \right) \begin{pmatrix}
A(K_R) \\ B(K_R)
\end{pmatrix} =0,
\ee
which implicitly defines the dispersion relation $E=E_I(k_y)$ of the interface modes.  
As for the boundary case \eqref{boundcond}, the two equations in Eq.~\eqref{interfacecond}
are nonlinear conditions for $k_y$ and $E$ which have to be solved simultaneously.
Depending on the parameter values, our numerical analysis shows that such solutions indeed exist. 
Typical results for the dispersion relation $E_I(k_y)$ are shown
in Fig.~\ref{fig9}.
We find interface modes of type I-II or type II-II, where the latter modes 
can only exist for $Md>2$. For the parameters in Fig.~\ref{fig9}, 
there are no type I-I interface modes. 
In fact, the absence of type I-I modes is a generic feature 
which can be rationalized by observing that their dispersion should 
originate from one of the two crossing points $(k_y=0,E=\pm V_s)$, 
but at the same time it should satisfy the conditions 
$k_y^2>(E_I\pm V_s)^2$. Clearly, both requirements are incompatible.

In analogy to the boundary modes in Sec.~\ref{sec4}, we expect such
interface modes to affect transport properties.  In addition, 
they should be observable by STM or tunneling spectroscopy.

\section{Conclusions}\label{sec6}

Our analysis of 2D Dirac fermions in a piecewise-constant mass superlattice, where 
the mass term periodically changes sign, shows a remarkable richness.
We have shown that
the low-energy part of the spectrum is spanned by the chiral zero modes tied to the
zero-mass lines of the superlattice.  Apart from the resulting anisotropic Dirac cone
dispersion, we also predict nontrivial boundary modes as well as interface modes near potential
steps.  Those modes exist in two different types.  Type-I modes require a momentum $|k_y|$
parallel to the zero-mass lines which is larger than the energy $|E|$.  
Instead, type-II modes emerge at small $|k_y|$ but exist only for $Md>2$, 
where $M$ is the amplitude of the mass term and $d$ the superlattice period.  
Both types of evanescent states could affect transport properties and 
should be observable by STM techniques. 
 
Although our results have been derived for a particular exactly solvable model, 
we have also shown that in the regime $Md\gg 1$, the low-energy physics is directly connected 
to the chiral zero modes localized at the zero-mass lines, and therefore is generic to all Dirac 
mass superlattices where the mass alternates between positive and negative values,
including periodic arrays of topological junctions between Chern insulators with different Chern numbers.
 
The low-energy theory put forward in this work points to several interesting extensions. 
First, the inclusion of an orbital magnetic field along the $z$-direction allows one to 
study the interplay of  Landau level formation and quantum Hall physics with the phenomena 
discussed above. Second, since we have a model of coupled 1D chiral fermions, 
bosonization methods \cite{Gogolin1998} can be used to construct solvable nonperturbative theories 
of this 2D system in the presence of electron-electron interactions.  

Zero-line modes similar to those discussed in our work have also been reported in recent 
experiments performed on magnetic topological insulators \cite{Zhao2023} which realize interfaces 
between quantum anomalous Hall insulators \cite{QAHreview} with different Chern numbers.  
We expect that our results will also be relevant in this platform.
Theoretical predictions for layer-dependent zero-line modes in antiferromagnetic topological insulator 
multilayer structures based on MnBi$_2$Te$_4$ \cite{Liang2023} suggest that our theory can also be applied in that context. 
An important caveat when comparing our results to experiments concerns the idealized step-like mass term 
considered here.  While this simplification allowed us to obtain exact analytical solutions, 
for smooth mass kinks, additional states localized at the kinks can emerge at elevated energies,  
so-called  Volkov-Pankratov states \cite{zarenia2012,Goerbig2017,tineke2020}.  
However, such states are non-chiral and are expected to 
cause distinct transport and spectroscopical features than the chiral states discussed in our work. 

To conclude, we hope that the results put forward here will inspire future experimental 
and theoretical work along these lines.

\begin{acknowledgments}
We thank Dario Bercioux and Alex Zazunov for valuable discussions.  
An extended stay of A.~Miserocchi at HHU D\"usseldorf has
been funded by the Erasmus+ for Traineeships program and by the University of Padova.
We acknowledge funding by the Deutsche Forschungsgemeinschaft (DFG, German Research Foundation),
Projektnummer 277101999 - TRR 183 (project B04), Normalverfahren Projektnummer EG 96-13/1, 
and under Germany's Excellence Strategy - Cluster of Excellence Matter and Light for 
Quantum Computing (ML4Q) EXC 2004/1 - 390534769.
\end{acknowledgments}

\appendix
\section{Matrix properties}\label{appA}

We here summarize useful algebraic relations involving the matrix 
$W_{M}(x)$ in Eq.~\eqref{WM}. We first note that its inverse is given by
\be\label{WMinv}
W_{M}^{-1}(x) = \frac{1}{2\kappa}\begin{pmatrix}
(k_y+\kappa)e^{-\kappa x} & i(M+E)e^{-\kappa x} \\
-(k_y-\kappa) e^{\kappa x}  & -i(M+E)e^{\kappa x}  
\end{pmatrix}
\ee 
with $\kappa$ in Eq.~\eqref{kappadef}.
Second, we observe that the determinant of $W_M(x)$ is $x$-independent,
$\det W_{M}^{}(x) = \frac{2i\kappa}{M+E}$. 
Third,  Eqs.~\eqref{WM} and \eqref{WMinv} imply  the relation
\begin{widetext}
\be\label{aux1}
W^{-1}_{-M}(x) \, W_{M}^{}(x) = 
\frac{1}{\kappa(E+M)}\begin{pmatrix}
E\kappa +Mk_y & e^{-2\kappa x}(\kappa+k_y)M \\ 
e^{2\kappa x}(\kappa-k_y)M & E\kappa - Mk_y
\end{pmatrix}.
\ee
Fourth, for real $\kappa$ corresponding to $E^2<k_y^2+M^2$, 
we find
\bea \label{jxx1}
W^\dagger_{M}(x) \, W_{M}(x) &=&  
\begin{pmatrix}
e^{2\kappa x}\left( 1+ \left(\frac{k_y-\kappa}{E+M}\right)^2\right)  & \frac{2E}{E+M} \\ 
 \frac{2E}{E+M} & e^{-2\kappa x}\left( 1+ \left(\frac{k_y+\kappa}{E+M} \right)^2\right)
\end{pmatrix},\\ W^\dagger_{M}(x) \, \sigma_x \, W_{M}(x) &=&  
-\frac{2\kappa}{E+M}\sigma_y, \quad
W^\dagger_{M}(x) \, \sigma_y \, W_{M}(x) =  \frac{2}{E+M} 
\begin{pmatrix} 
e^{2 \kappa x} (k_y-\kappa) & k_y \\
 k_y &  e^{-2 \kappa x} (k_y+\kappa)
  \end{pmatrix}.\nonumber
\eea
For $\kappa=ik$ with real $k>0$, we instead find
\bea\label{jxx2}
W^\dagger_{M}(x) \, W_{M}^{}(x) &=& 
\begin{pmatrix}
\frac{2E}{E+M} & e^{-2ik x}\left( 1 - \left( \frac{-k+ik_y}{E+M} \right)^2 \right)   \\ 
 e^{2ik  x}\left( 1 - \left( \frac{k+ik_y}{E+M} \right)^2 \right) &  \frac{2E}{E+M}
\end{pmatrix}, \\  \nonumber
W^\dagger_{M}(x) \, \sigma_x \, W_{M}^{}(x) &=&  
\frac{2k}{E+M}\sigma_z ,\quad
W^\dagger_{M}(x) \, \sigma_y \, W_{M}^{}(x) = \frac{2}{E+M} 
\begin{pmatrix} 
k_y &e^{-2 i k x} ( k_y+ik) ,  \\
e^{2 i k x} (k_y-ik) & k_y
 \end{pmatrix}.
\eea
Next, the matrix $\Omega_B$ in Eq.~\eqref{OmegaB} for the mass-barrier problem in Sec.~\ref{sec2c}, 
is given by
\be\label{OmegaBA}
\Omega_B= \frac{1}{\kappa^2(E^2-M^2)}
\left(
\begin{array}{cc}
(E^2-M^2)[k_y^2 - E^2 + M^2  e^{-2\kappa \ell}] \quad 
& \quad -2M(\kappa +k_y)(E\kappa +k_yM)\sinh(\kappa\ell)
\vspace*{0.2cm} \\
2M(\kappa - k_y)(E\kappa -k_yM) \sinh(\kappa\ell)\quad 
& \quad (E^2-M^2)[k_y^2 -E^2 + M^2 e^{2\ell \kappa}]
\end{array}\right).
\ee
Similarly, the modified transfer matrix $\Omega$ in Eq.~\eqref{omega} for the 
periodic mass profile \eqref{periodicmass} reads
\be\label{Omegaexplicit}
\Omega  =\frac{1}{\kappa^2(E^2-M^2)} \left(
\begin{array}{cc} (E^2-M^2) [M^2+(k_y^2-E^2)e^{\kappa d}] \quad 
& \quad M (1-e^{-\kappa d}) (\kappa+k_y)(E\kappa-Mk_y) \vspace*{0.2cm}\\ 
 M (1-e^{\kappa d})(\kappa-k_y)(E\kappa+Mk_y) \quad 
& \quad (E^2-M^2) [M^2+(k_y^2-E^2)e^{-\kappa d}]
\end{array}\right).
\ee
For $E^2<k_y^2+M^2$ such that $\kappa$ is real, the matrix elements of $\Omega$ are also real. 
For completeness, we also specify the elements of the symmetric transfer matrix $T$:
\bea
\nonumber
T_{11} & =& \frac{M^2+(k_y^2-E^2)\cosh(\kappa d)}{\kappa^2} + 
\frac{ME[\cosh(\kappa d)-1]+ k_y\kappa \sinh(\kappa d)}{\kappa^2}, \\
T_{12}&=&T_{21} \label{Texplicit}
 = i\,\frac{E\kappa \sinh(\kappa d)+Mk_y(\cosh(\kappa d )-1)}{\kappa^2}, \\
 \nonumber
T_{22} & =&\frac{M^2+(k_y^2-E^2)\cosh(\kappa d)}{\kappa^2} - 
\frac{ME[\cosh(\kappa d )-1]+ k_y\kappa \sinh( \kappa d)}{\kappa^2}.
\eea
\end{widetext}

\bibliographystyle{apsrev4-1}
\bibliography{biblio}

\end{document}